\pdfoutput=1
\documentclass[a4paper,symmetric,justified]{tufte-book} 

\usepackage[utf8]{inputenc}
\usepackage{microtype} 

\usepackage{lipsum} 

\usepackage{booktabs} 

\usepackage{braket,amsmath} 

\usepackage{graphicx} 
\graphicspath{{graphics/}} 
\setkeys{Gin}{width=\linewidth,totalheight=\textheight,keepaspectratio} 

\usepackage{fancyvrb} 
\fvset{fontsize=\normalsize} 

\usepackage{xspace} 

\newcommand{\monthyear}{\ifcase\month\or January\or February\or March\or April\or May\or June\or July\or August\or September\or October\or November\or December\fi\space\number\year} 

\newcommand*\mean[2]{\ensuremath\left\langle{#1}\right\rangle_{#2}}
\DeclareMathOperator{\Tr}{Tr}

\DeclareMathOperator{\conv}{conv}

\usepackage{etoolbox}
\usepackage{makeidx} 
\makeindex 

\title{Uncertainty relations \\and joint numerical ranges} 
\subtitle{MSc thesis under the supervision of Prof. Karol Życzkowski}
\author{Konrad Szymański}

\publisher{Marian Smoluchowski Institute of Physics, \\ \mbox{Uniwersytet Jagielloński}\\ {\small Kraków, July 2017}} 

\begin{document}

\frontmatter

\maketitle 


\tableofcontents 


\cleardoublepage
~\vfill
\begin{doublespace}
\noindent\fontsize{18}{22}\selectfont\itshape
\nohyphenation
To MS, AS, unknow, DS, JJK, AM, FF, KŻ, SM, P, N.
\thispagestyle{empty}
\end{doublespace}
\vfill
\vfill


\cleardoublepage


\chapter*{Introduction} 

Noncommutativity lies at the heart of quantum theory and provides a rich set of mathematical and physical questions. In this work, I address this topic through the concept of Joint Numerical Range (JNR) --- the set of simultaneously attainable expectation values of multiple quantum observables, which in general do not commute. 

The thesis is divided into several chapters:

\section{Quantum states} 
Geometry of the set of quantum states of size $d$ is closely related to JNR, which provides a nice tool to analyze the former set. The problem of determining the intricate structure of this set is known to quickly become hard as the dimensionality grows (approximately as $d^2$). The difficulty stems from the nonlinear constraints put on the set of parameters.

In this chapter I briefly introduce the formalism of \emph{density matrices}, which will prove useful in the later sections.


\section{Joint Numerical Range}
Joint Numerical Range of a collection of $k$ operators, or JNR for short, is an object capturing the notion of simultaneous measurement of {\emph{averages}} --- expectation values of multiple observables. This is precisely the set of values which are simultaneously attainable for fixed observables $(F_1,F_2,\ldots,F_k)$ over a given quantum state $\rho\in\mathcal{M}_d$: if we had a function taking quantum state and returning the tuple of average values:
\begin{equation}
\mathbb{E}(\rho)=(\mean{F_1}{\rho},\mean{F_1}{\rho},\ldots,\mean{F_k}{\rho}),
\end{equation}
Joint Numerical Range $L(F_1,\ldots,F_k)$ is precisely $\mathbb{E}[\mathcal{M}_d]$.
This chapter explains further the definition, comments on the basic properties of JNR and presents results existing in the literature so far.

\section{Phase transitions} 
It turns out that from the JNR of operators determined by analyzed Hamiltonian it is possible to deduce properties of a quantum system at zero temperature. In particular one can identify and investigate the phase transitions the system undergoes as the parameters of Hamiltonian vary.

This chapter explains the connection between JNR and phase transitions at zero temperature. 

\section{Uncertainty relations}
Uncertainty relations, providing a numerical formalization of the indeterminacy of quantum measurement, are deeply linked with notion of the Joint Numerical Range of selected observables.
In this chapter we provide analytical and numerical tools to develop the theorems and new bounds for uncertainty relations.

\chapter*{Notation}
\begin{itemize}
	\item $\mathcal{H}_d$ -- $d$-dimensional complex Hilbert space,
	\item $\mathcal{M}_d$ -- set of density matrices acting on $d$--dimensional Hilbert space,
	\item $\sigma(X)$ -- set of eigenvalues of an operator $X$,
	\item $P^{(X)}_\lambda$ -- projector onto the eigenspace of $X$ corresponding to the eigenvalue $\lambda$.
	\item $\mathbb{E}^{F_1,\ldots,F_k}$ --- the `measure average' map, $\mathbb{E}(\rho)=(\mean{F_1}{\rho},\ldots,\mean{F_k}{\rho})$. Which exactly observables $(F_1,\ldots,F_k)$ are measured is usually clear from the context -- in such cases, the list in the superscript is omitted.
	\item $[\psi]$ --- a projector onto $\ket{\psi}$, Dirac dyad $\ket{\psi}\bra{\psi}$. Occasionally used over sets, in which case threads over all elements, $[\{\psi,\phi\}]=\{\ket{\psi}\bra{\psi},\ket{\phi}\bra{\phi}\}$
	\item $\conv S$ --- convex hull of $S$, i.e. $\left\{\sum \alpha_i p_i | \sum \alpha_i=1, \alpha_i>0, p_i \in S\right\}$ 
	\item $L(F_1,\ldots,F_k)$ -- Joint Numerical Range of operators $F_1,\ldots,F_k$, $\mathbb{E}^{F_1,\ldots,F_k}(\mathcal{M}_d)$
\end{itemize}



\mainmatter


\chapter{Quantum states}
\label{ch:1}


\begin{fullwidth}
In this introductory section I explain the notion of an \emph{density operator}, which is a generalization of quantum state which allows for \emph{statistical ensembles}. This generalization will prove useful in later sections --- not only it simplifies several calculations, but also provides a better view on the nature of quantum entanglement.
\end{fullwidth}

\section{Pure states}

We concentrate on a finite--dimensional Hilbert spaces only. Let the $d$-dimensional Hilbert space over complex field ${C}$ be denoted by $\mathcal{H}_d$. Every realization of $\mathcal{H}_d$ is isomorphic to ${C}^d$, therefore it is natural to identify tuples

\begin{equation*}
\ket{\psi}=\begin{pmatrix}\psi_1,\ldots,\psi_d\end{pmatrix}^T
\end{equation*}

with vectors in $\mathcal{H}_d$. The identification goes further: in physical systems, the vectors $\ket{\psi}$ and $e^{i \phi} \ket{\psi}$ are indistinguishable.
\subsection{Bloch sphere}

A simple example of nontrivial Hilbert space is the two--dimensional complex case $\mathcal{H}_2$, describing a qubit. Since on a general vector
\begin{equation}
\begin{pmatrix} \psi_0 \\ \psi_1 \end{pmatrix}
\end{equation}
we impose the normalization condition $|\psi_0|^2 + |\psi_1|^2 = 1 $, we can rewrite it as
\begin{equation}
\label{eqn:expr3}
\begin{pmatrix} e^{i \alpha_0} r \\ e^{i \alpha_1} \sqrt{1-r^2} \end{pmatrix},
\end{equation}
where $\alpha_0, \alpha_1, r \in {R}$. Furthermore, since we identified the vectors differing only by overall phase, expression \eqref{eqn:expr3} is equivalent to the vector
\begin{equation}
\begin{pmatrix}  r \\ e^{i (\alpha_1-\alpha_0)} \sqrt{1-r^2} \end{pmatrix}.
\end{equation}
It is now natural to rewrite this vector as

\begin{equation}
\ket{\theta,\phi}=\begin{pmatrix}  \cos \frac\theta 2 \\ e^{i \phi} \sin \frac\theta 2 \end{pmatrix}.
\end{equation}

This form clearly shows the possible geometrical representation of one pure qubit state: as a point on a \emph{Bloch sphere}. 

\section{Density operators}
Pure states do not account for all of the quantum phenomena. Often when considering a state belonging to the composite Hilbert space $\mathcal{H}_A\otimes\mathcal{H}_B$ and performing multiple observations in one of the subspaces -- say, $\mathcal{H}_A$ -- only, the results are incompatible with \emph{any} local state vector\cite{bell1964einstein}. An extension is needed, and the simplest one is the formalism of density operators.
One of the possible motivations is the need for considering \emph{statistical mixtures} of pure states and their evolution. This may stem from experiment: imagine a device capable of generating states with certain (classical) probabilities: so the output state of a device is a set of tuples $($state$,$probability$)$:
\begin{equation}
(\Psi)=\{(\ket{\psi_1},p_1),\ldots,(\ket{\psi_n},p_n)\},
\end{equation}
where ${p_i}$ obey usual constraints: $p_i\ge0, \sum\limits_{i=1}^n p_i=1$. The average of any quantum observable $A$ is the weighted sum of averages over constituent states:
\begin{equation}
\mean{A}{(\Psi)}=\sum_{i=1}^n p_i \mean{A}{\ket{\psi_i}}=\sum_{i=1}^n p_i \braket{\psi_i | A \psi_i}.
\end{equation}
From the purely mathematical standpoint, the rightmost expression may be rewritten, using the cyclic property of trace, as
\begin{equation}
\mean{A}{(\Psi)}=\sum_{i=1}^n p_i \Tr [\ket{\psi_i}\bra{\psi_i} A] =\Tr \left[\left(\sum_{i=1}^n p_i \ket{\psi_i}\bra{\psi_i} \right) A\right].
\end{equation}
Observe that it is possible to write any quantum average as $\Tr \rho A$ -- where $\rho$ does not depend on the observable $A$ at all. It is composed from the projectors onto the constituents and associated probabilities \emph{only}. This object,
\begin{equation}
\rho_{(\Psi)}=\sum_{i=1}^n p_i \ket{\psi_i}\bra{\psi_i},
\end{equation} 
is called a \emph{density matrix}. Now, of course, we have lost something -- many different ensembles $(\Psi)$ may correspond to single density matrix\footnote{The simplest example is $\{(\ket{0},\frac12),(\ket{1},\frac12)\}$ and $\{(\ket{+},\frac12),(\ket{-},\frac12)\}$ -- the resulting $\rho$ is for both cases $\mathbb{1}/2$}, but the omitted part was not even potentially observable to begin with.

The notion of density matrices plays a significant role on its own and it is not only limited to the randomly prepared systems. Consider a state $\ket{\psi}$ in the composite Hilbert space $\mathcal{H}_A\otimes\mathcal{H}_B$, which components in a basis are indexed according to the basis in the subsystems
\begin{equation}
\psi_{i,j}=\braket{e^{(A)}_i\otimes e^{(B)}_j|\psi}.
\end{equation}
Now if we want to measure an average of any local operator acting on the first subspace, $A\otimes \mathbb{1}$, we again can refer to the density matrices. By simple calculation,
\begin{equation}
\mean{A\otimes\mathbb{1}}{\ket{\psi}}=\sum \psi^*_{i,j} A_{i,i'} \delta_{j,j'} \psi_{i',j'}= \sum \psi^*_{i,j}\psi_{i',j} A_{i,i'},
\end{equation}
we have arrived at the fact that for any intents and purposes of measurement a state, if measured only in one subsystem, is effectively described by the \emph{reduced} or \emph{partially traced} density matrix:
\begin{equation}
\mean{A\otimes\mathbb{1}}{\ket{\psi}} = \Tr \left [(\Tr_B \ket{\psi}\bra{\psi}) A \right],
\end{equation}
where, once again, $\rho_A=\Tr_B \ket{\psi}\bra{\psi}$ is independent on $A$:
\begin{equation}
\rho_A=\Tr_B \ket{\psi}\bra{\psi}=\left(\sum \psi^*_{i,j} \psi_{i',j}\right)_{i,i'}.
\end{equation}

\subsection{Properties}
Density operators have several important properties, which are extensively used:
\begin{enumerate}
	\item any $\rho$, being a convex combination of \emph{positive semidefinite}\footnote{i.e. having all eigenvalues nonnegative} projectors, it itself positive semidefinite,
	\item analogously, density operators are \emph{Hermitian},
	\item $\Tr \rho = 1$, owing to the normalization of probability,
	\item for any $\rho, \rho'$ being density operators and $\alpha\in[0,1]$, $\alpha \rho+(1-\alpha)\rho'$ is a valid density operator. The set of all density operators is therefore \emph{convex}.
\end{enumerate}
We denote the set of all density operators of fixed rank $d$ as $\mathcal{M}_d$
\subsection{Bloch ball}
\begin{marginfigure}
\includegraphics{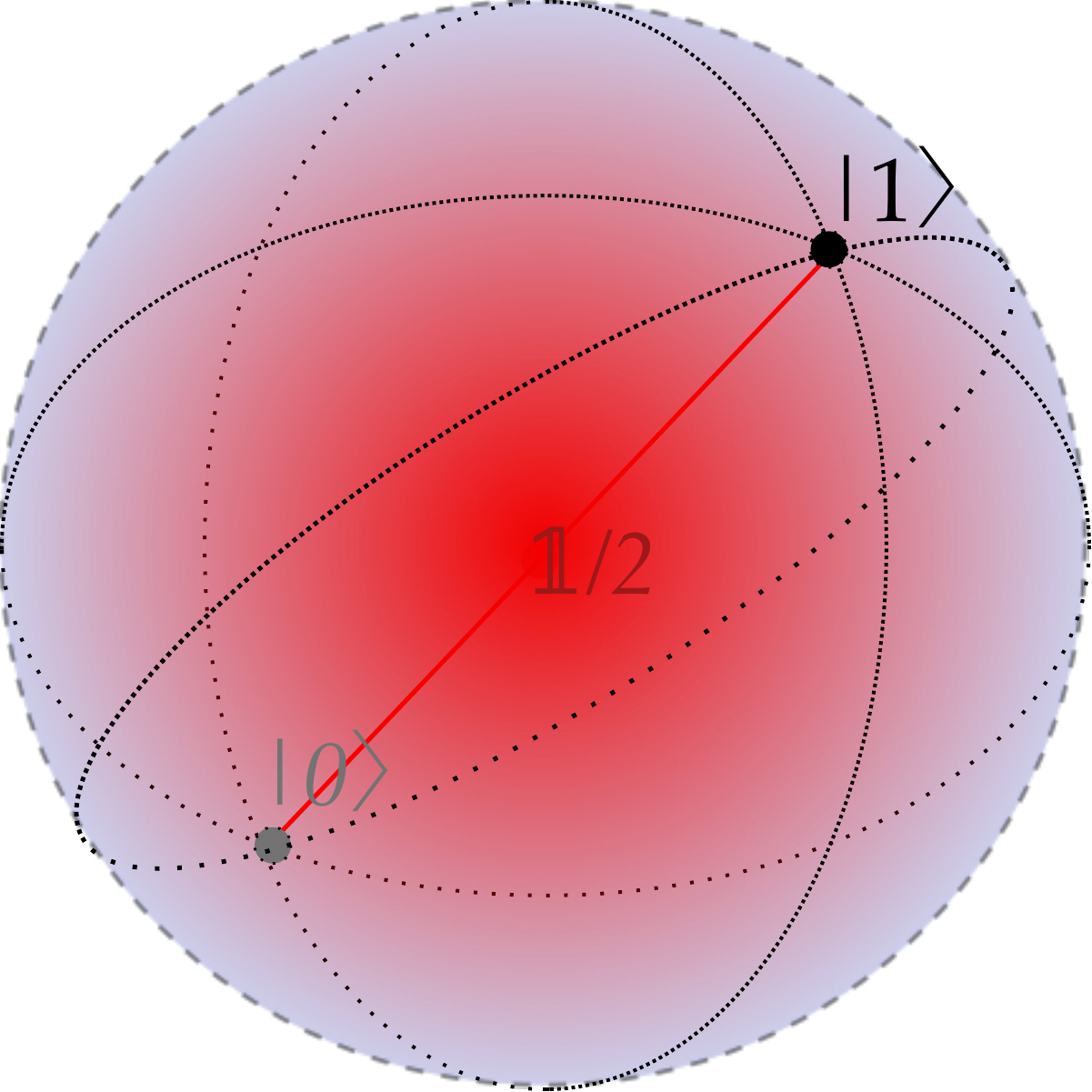}
\caption{Visual depiction of the Bloch ball. Pure states $\ket{0},~\ket{1}$ as well as completely mixed state are shown.}
\end{marginfigure}
To complete the analysis of a single qubit let us extend the visual representation of such a state -- a point on a \emph{sphere} -- to the mixed states. Since every unit trace, Hermitian matrix may be parametrized like
\begin{equation}
\rho=\begin{pmatrix}\frac12-z&x+i y\\x-i y& \frac12+z\end{pmatrix}
\end{equation}
with $x,y,z\in\mathbb{R}$, we are left with the condition for the $\rho$ to be positive semidefinite. If the eigenvalues are $\lambda_-,\lambda_+$ and $\lambda_++\lambda_-=1$, as required by unit trace, it turns out that $\det \rho=\lambda_- \lambda_+$ is positive if and only if $\rho$ is positive, and since the determinant has a simple form
\begin{equation}
\det\rho=\frac14-(x^2+y^2+z^2),
\end{equation}
The set of density matrices, embedded in the unit trace Hermitian space, forms an Euclidean ball\footnote{In the sense of Hilbert-Schmidt distance.} centered at $\begin{pmatrix}\frac12&0\\0&\frac12\end{pmatrix}$. 
The pure states form the boundary of this ball -- the Bloch \emph{sphere}.


\chapter{Joint Numerical Range}
\label{ch:2}
\begin{fullwidth}
In this chapter we present the theory of numerical ranges -- intuitively, the set of simultaneously attainable quantum expectation values of selected observables. As in general the observables need not commute, the resulting objects may have nontrivial shapes. In the end, I will present some of the possible generalizations of JNR.
\end{fullwidth}
\section{History}
First articles in the theory of numerical ranges date as early as 1918\cite{toeplitz1918algebraische}\cite{hausdorff1919}. The theory back in the time was mostly of pure mathematical interest, so it was natural for the founders to consider the numerical range of arbitrary, not necessarily self-conjugate bilinear form $F$, defined (in Dirac notation) as
\begin{equation}
W(F)=\left\{\frac{\braket{x|F x}}{\braket{x|x}}~:~\ket{x} \in \mathbb{C}^n \textbackslash \{0\} \right\}.
\end{equation}
The first important result was the \emph{Hausdorff-Toeplitz theorem}, stating that any $W(F)$ is \emph{convex} and \emph{compact}\footnote{As it will be shown, this is not true if larger number of observables is considered.}
\section{Joint Numerical Range}
Joint Numerical Range (or JNR), an extension of the notion of numerical range, is the central object of this thesis. There are important differences:
\begin{enumerate}
	\item an arbitrary number of observables are simultaneously measured,\footnote{This enables restriction to -- more natural for a physicist -- Hermitian operators only, as every nonhermitian $F$ may be equivalently transformed into its Hermitian $F_H$ and antihermitian $F_{A}$ part, such that $F=F_H+i F_{A}$.}
	\item the set over which the `measure average' $\mathbb{E}$ map acts on is extended to the entire set of density matrices.
\end{enumerate}
These extensions are roughly resembled by
\begin{equation}
\braket{x|F x} \mapsto (\Tr \rho F_1, \Tr \rho F_2, \ldots, \Tr \rho F_k).
\end{equation}
The function taking as an argument a state and returning tuple of quantum averages is denoted by $\mathbb{E}$:
\begin{equation}
\mathbb{E}^{F_1,\ldots, F_k}(\rho)=(\Tr \rho F_1, \Tr \rho F_2, \ldots, \Tr \rho F_k).
\end{equation}
To distinguish Joint Numerical Range from the numerical range of Hausdorff and Toeplitz we use a different letter to denote it -- $L$ instead of $W$. The full definition of JNR is therefore
\begin{equation}
L(F_1,\ldots,F_k)=\{ \mathbb{E}^{F_1,\ldots,F_k} (\rho) : \rho \in \mathcal{M}_d \}.
\end{equation}
~\\
The second point -- mapping over whole state space -- needs clarification: it is in fact needed to ensure that the resulting object is \emph{convex}\footnote{i.e. for every pair $x,y$ contained in the set and $\alpha\in[0,1]$, the point $\alpha x+ (1-\alpha)y$ belongs to the set.}. As a simple example, if we consider the numerical range restricted to pure \emph{qubit} states with three Pauli operators,
\begin{equation}
\sigma_x=\begin{pmatrix}0&1\\1&0\end{pmatrix},~\sigma_y=\begin{pmatrix}0&i\\-i&0\end{pmatrix},~\sigma_z=\begin{pmatrix}1&0\\0&-1\end{pmatrix},
\end{equation}
the resulting object is a \emph{hollow} sphere $S^2$, as for the pure states the following holds:
\begin{equation}
\mean{\sigma_x}{\ket{\psi}}^2+\mean{\sigma_y}{\ket{\psi}}^2+\mean{\sigma_z}{\ket{\psi}}^2=\frac14.
\end{equation}
Since the `measure-averages' map $\mathbb{E}$ is linear and in the case of JNR acts on a convex set, the result -- which we may interpret as a projection of the entire set of quantum states\footnote{Or rather, as an affine transformation of such a projection.} onto a $k$-dimensional subspace.
\section{Determination of the boundary}
As any convex object is the convex hull of its extremal points, only knowledge of the latter set is practically needed. Here I will present the method of generating the entire boundary of JNR of 2 operators and extend it to an arbitrary number of operators $k$.
\subsection{2 operators}
The method dates back to the early works
 It takes advantage of a simple observation: the set of rightmost points\footnote{In the case of 2 operators it is either a singleton or a vertical segment.} has $x$-coordinate $\lambda_{\text{max}}(X)$ --- the maximum eigenvalue of $X$. Of course, its preimage is given\footnote{projector onto eigenspace of $\lambda_{max}$ and their convex combinations} by $\conv [\Lambda(X,\lambda_{max})]$. We can map this set over the second operator $Y$ obtaining a point or a line -- in any case, this image forms a part of the boundary of the JNR.
Knowing a technique to identify the rightmost points of JNR of a given collection of $k$ operators one can generalize it to describe the entire boundary of JNR.%
 Observe that the JNR of the operators  `rotated' by $\theta$:
\begin{equation}
L_\theta (X,Y) =L(X \cos\theta + Y \sin\theta, Y \cos\theta - X \sin\theta)
\end{equation}
give \emph{the rotated JNR} $L(X,Y)$. Since we can calculate the set of rightmost points of $L_\theta(X,Y)$ --- image of eigenspace to maximum eigenvalue of $X \cos\theta + Y \sin\theta$ and map it over $\mathbb{E}^{(X,Y)}$ and we know that the resulting set is in the boundary of $L(X,Y)$.

\newthought{The numerical procedure} for computing the boundary of JNR of $X,Y$ therefore looks like:
\begin{itemize}
	\item for $\theta$ in $[0,2\pi]$:
	\item calculate $\lambda_{max}(\cos \theta X + \sin\theta Y)$ and projector onto the eigenspace $P$
	\item append the image of the eigenspace $P$ under $\mean{X}{},\mean{Y}{}$ to the boundary.
\end{itemize}
\subsection{Arbitrary number of operators}
The above procedure can be generalized to an arbitrary number of operators $k$ of an arbitrary dimension $d$ with only minor modifications. The method is based on the convexity of JNR, which can be determined by supporting hypersurfaces.
\begin{itemize}
	\item for any $\vec n$ in $S^{n-1}$:
	\item calculate $\lambda_{max}(\sum n_i F_i)$ and projector onto the eigenspace $P$
	\item calculate $L(P F_1 P,\ldots,P F_k P)$ and append it to the boundary.
\end{itemize}
\section{Classification of JNRs}
The problem of classifying the JNR based on the features of their boundaries proved to be hard to tackle. Only few cases have been analyzed completely: 2 and 3 operators acting on a qubit and qutrit. Since the qubit case was studied long time ago \footnote{The $\mathbb{E}$ map is linear and acts on a Bloch \emph{ball}, thus JNR is an ellipse or ellipsoid, possibly degenerated to a line.}, the qutrit case is discussed further.
\subsection{2 operators acting on a qutrit}
\begin{figure}
\label{fig:clask2}
0. \raisebox{-2cm}{\includegraphics[width=0.4\linewidth]{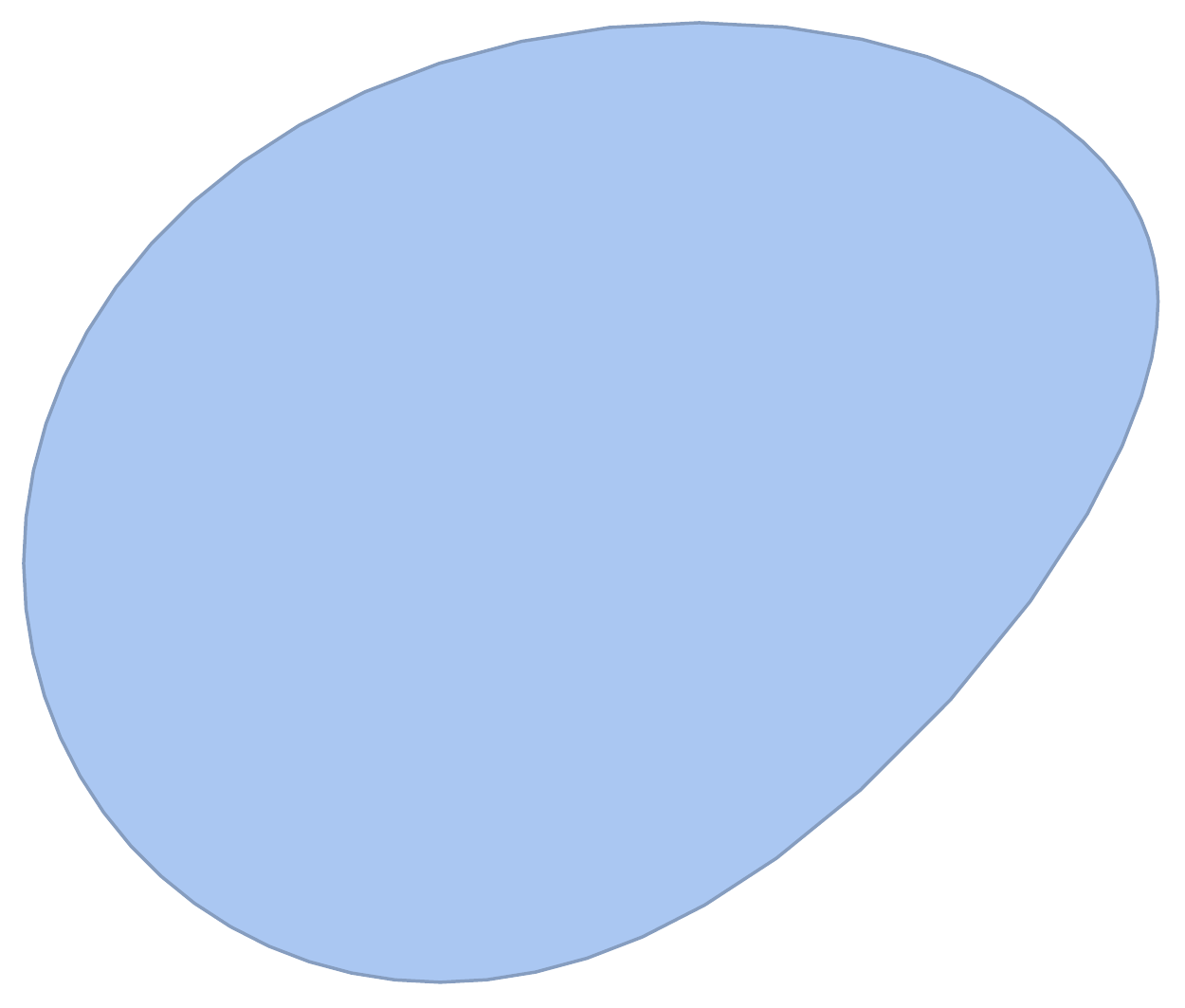}}%
1. \raisebox{-2cm}{\includegraphics[width=0.4\linewidth]{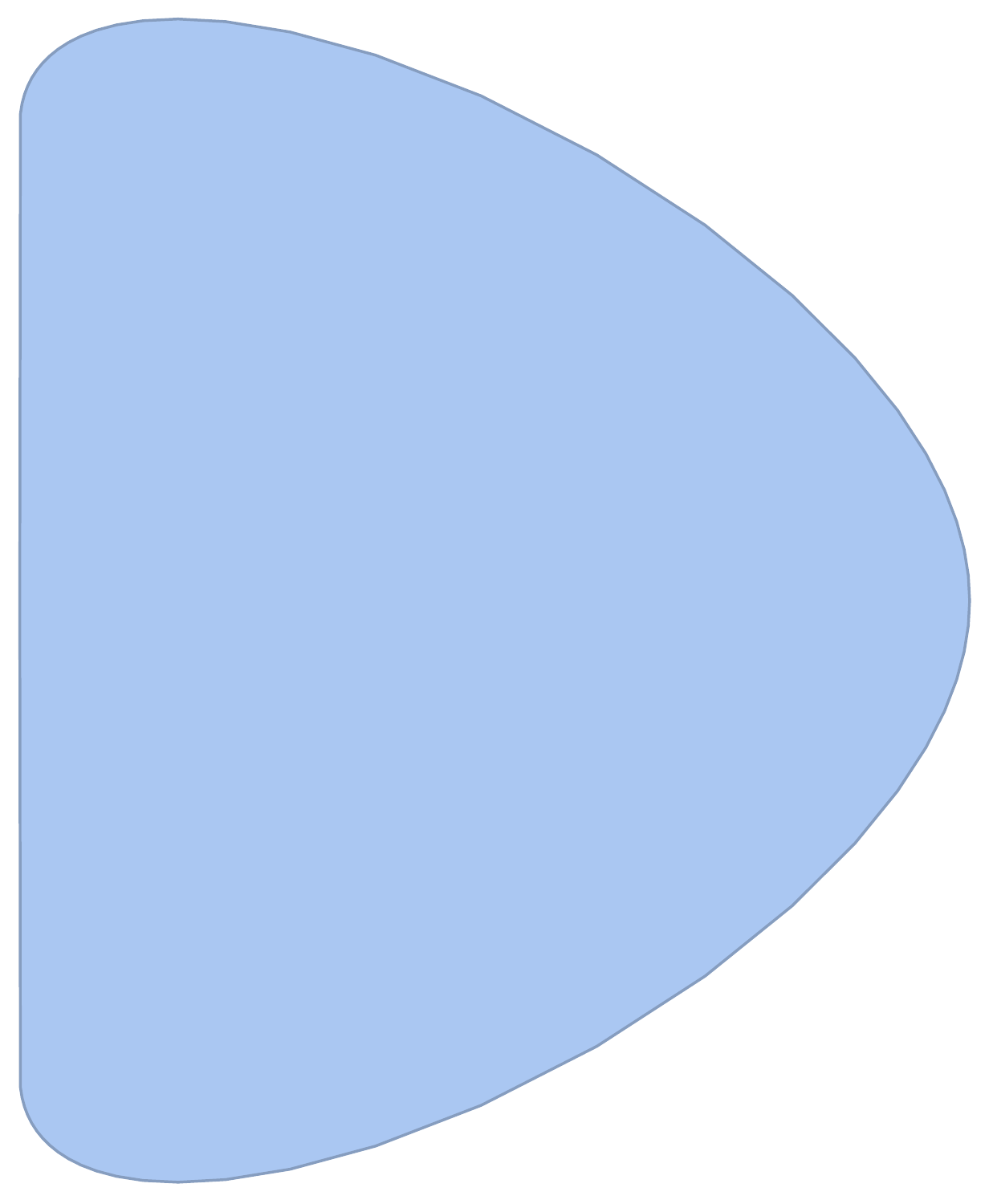}}\\%
2. \raisebox{-1cm}{\includegraphics[width=0.4\linewidth]{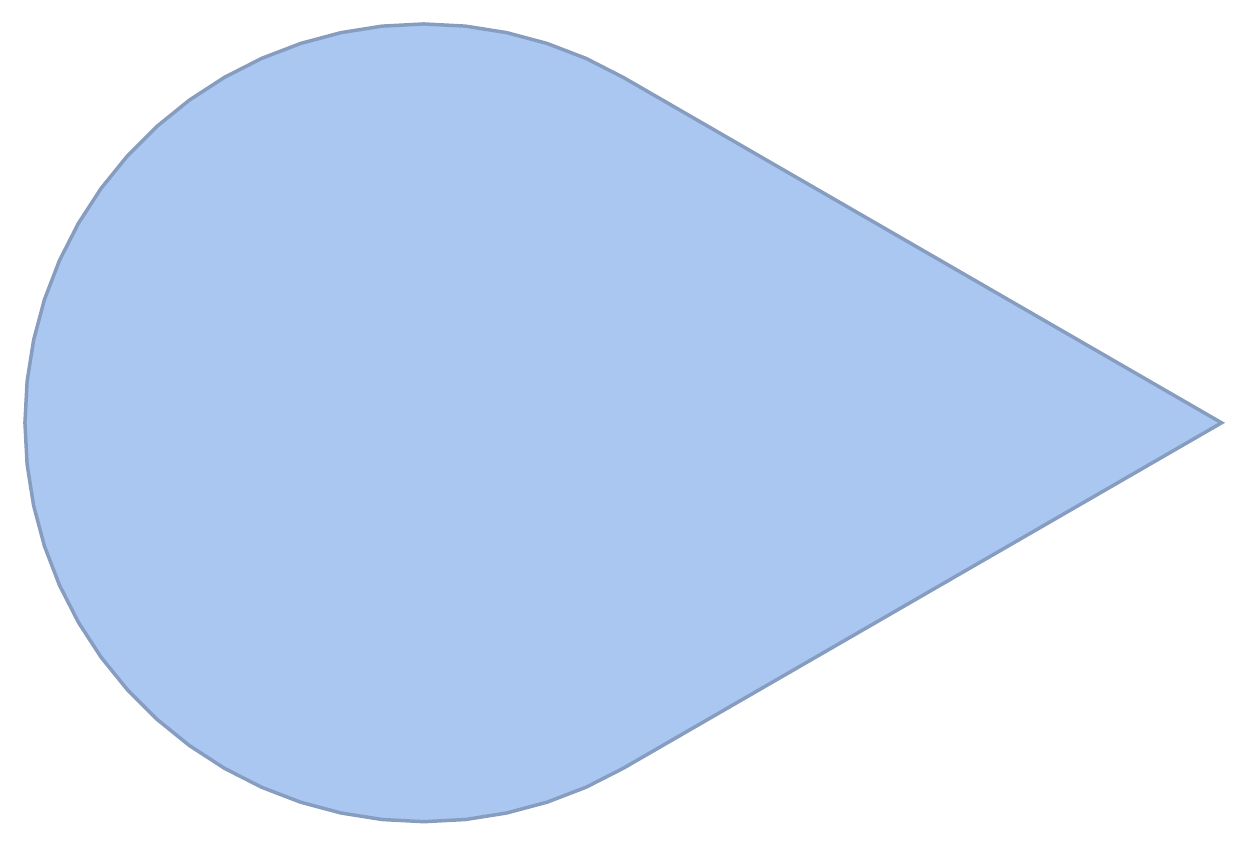}}%
3. \raisebox{-1cm}{\includegraphics[width=0.4\linewidth]{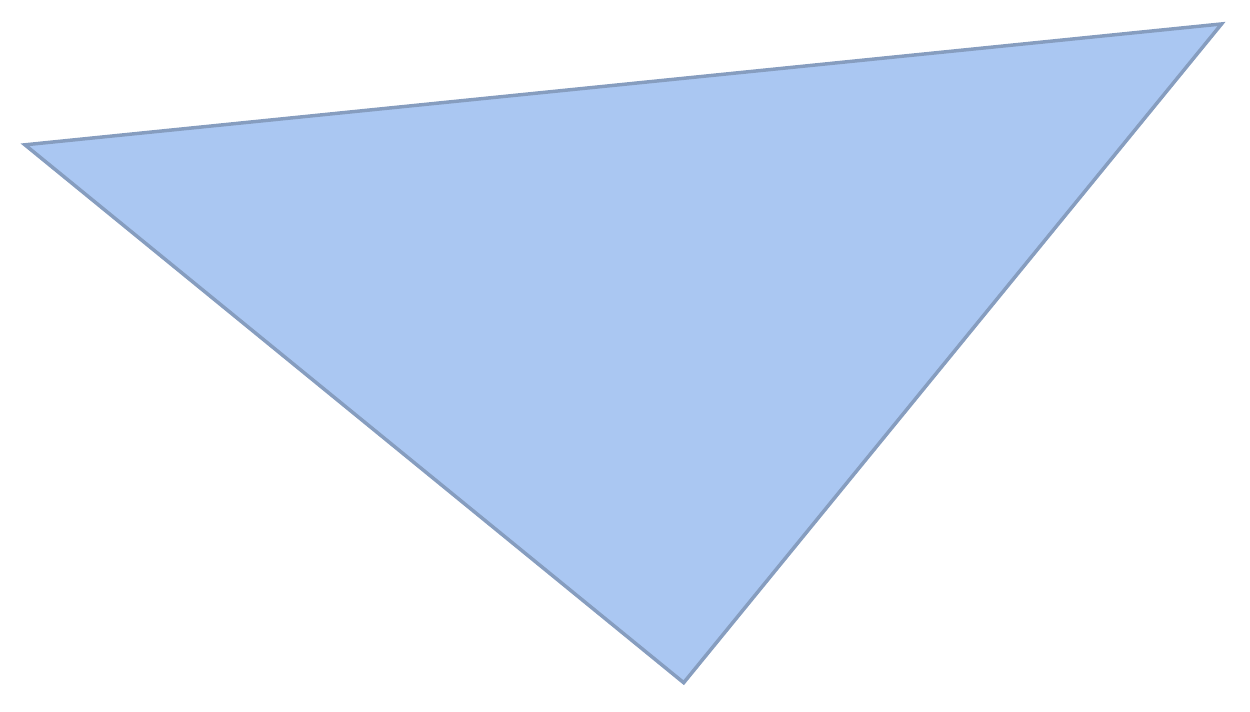}}
\caption{Four classes of JNR of two Hermitian operators of order three. The boundary contains 0, 1, 2, or 3 line segments.}
\end{figure}

The possible shapes of JNR of 2 operators of order $d=3$ are limited to four classes\cite{keeler1997numerical}:
\begin{itemize}
\item {an oval} -- object without any flat parts, with the boundary being a sextic curve,
\item {object with one flat part}, a convex hull of quatric curve,
\item convex hull of an ellipse and outside point, which has two connecting segment in the boundary
\item a triangle (emerging when the two operators commute), possibly degenerated.
\end{itemize}
See Fig. \ref{fig:clask2} for the examples of objects belonging to each class.
\subsection{3 operators acting on a qutrit}
Classification of JNR of 3 operators acting on a qutrit was carried out in 2016\cite{szymanski2016classification}. The main arguments for classification are:
\begin{enumerate}
	\item in this particular case, the restriction to pure states is possible -- $\mathbb{E}$ mapped over pure states only gives the same object in $\mathbb{R}^3$ as originally defined JNR,
	\item therefore, any flat part in the boundary is the image of Bloch sphere -- two-dimensional subspaces of $\mathcal{H}_3$,
	\item image under $\mathbb{E}$ of Bloch sphere is an \emph{ellipse}, possibly degenerated to a \emph{segment},
	\item two 2-D subspaces in $\mathcal{H}_3$ must share a common point, so all flat parts are mutually connected,
	\item convex geometry of $\mathbb{R}^3$ supports up to four mutually intersecting ellipses,
	\item if three ellipses are present in the boundary, $\mathbb{R}^3$ geometry does not allow for existence of any additional segment,
	\item if two segments are present in the boundary, there exist an infinite number of other segments.
\end{enumerate}
All configurations permitted by these rules are realized: there exist objects with
\begin{itemize}
	\item no flat parts in the boundary at all: $e=0, s=0$,\footnote{$e$ stands for number of ellipses in the boundary, $s$ for the number of segments.}
	\item one segment in the boundary, which we denote symbolically as \raisebox{-1.75mm}{\includegraphics[height=7mm]{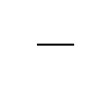}}\footnote{The classification symbols are taken from $^{\text{(5)}}$.}: $e=0, s=1$,
	\item one ellipse in the boundary, which we denote symbolically as \raisebox{-1.75mm}{\includegraphics[height=7mm]{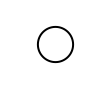}}: $e=1, s=0$,
	\item one ellipse and a segment, \raisebox{-1.75mm}{\includegraphics[height=7mm]{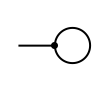}}, where the dot denotes the point of intersection: $e=1, s=1$,
	\item two ellipses in the boundary -- \raisebox{-1.75mm}{\includegraphics[height=7mm]{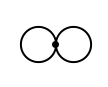}}: $e=2, s=0$,
	\item two ellipses and a segment -- \raisebox{-1.75mm}{\includegraphics[height=7mm]{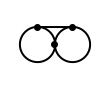}}: $e=2, s=1$,
	\item three ellipses -- \raisebox{-1.75mm}{\includegraphics[height=7mm]{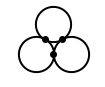}}: $e=3, s=0$,
	\item four ellipses -- \raisebox{-1.75mm}{\includegraphics[height=7mm]{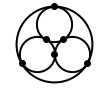}}: $e=4, s=0$.
\end{itemize}
This classification by the number of ellipses and segments may be shown in a concise manner:
\begin{figure}
\includegraphics[width=\linewidth]{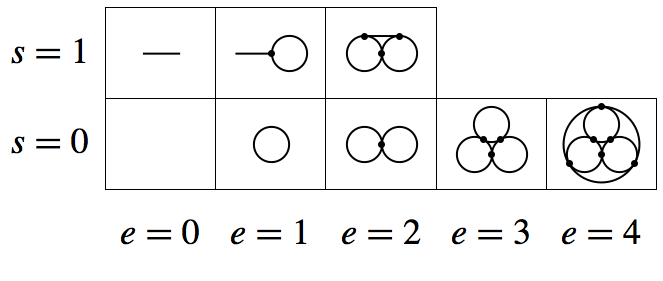}
\caption{Classification of JNR of three operators of rank three: $e$ stands for number of ellipses and $s$ number of segments. The circles denote elliptical flat parts in the boundary, segments correspond to segments, dots denote common points of two flat parts. Note that in the $e=4$ case there is outer circle.}
\end{figure}
\section{Possible generalizations}
Many possible generalizations of JNR have been discussed in previous works\cite{gustafson2012numerical}\cite{gawron2010restricted}. In this section I will present an important one, separable numerical range\footnote{Also known as \emph{product numerical range} and \emph{tensor numerical range}.} and original work generalizing the numerical range to non-zero temperatures.
\subsection{Separable numerical range}
Consider composite state space, $\mathcal{H}_d=\mathcal{H}_l\otimes\mathcal{H}_m$ (then $d=lm$). One of the possible generalization is restriction of the set $\mathbb{E}$ map acts on to some convex subset of $\mathcal{M}_d$. The example showing useful properties applicable in physics (but not only -- see \cite{friedland2014nuclear}) is the restriction to separable states, i.e., those of the form
\begin{equation}
\begin{split}
\rho=\sum_i p_i \ket{\alpha_i}\bra{\alpha_i} \otimes \ket{\beta_i}\bra{\beta_i},\\
p_i >0, ~~~ \sum_i p_i=1.
\end{split} 
\end{equation}
Such object, $L^{sep}(F_1,\ldots,F_n)$ may be compared with the larger set of $L(F_1,\ldots,F_n)$ and act as a nonlinear entanglement witness. If a quantum state is measured in a laboratory to give vector of expectation values $\vec f=(\mean{F_1}{},\ldots,\mean{F_n}{})\in L(F_1,\ldots,F_n)$ and it does not lie in $L^{sep}(F_1,\ldots,F_n)$ one may be infer that the original state is entangled. This object is additionally useful in studies of phase transitions in the approximation of infinite spatial dimensions\cite{chen2016physical}.

\newthought{Examples} of new behavior seen in separable numerical range are ubiquitous, even for the simplest, two-qubit system. Consider the state 
\begin{equation}
\ket{\psi_+}=\begin{pmatrix}1\\0\\0\\1\end{pmatrix},
\end{equation}
which is highly entangled\footnote{In fact, it is a maximally entangled state, i.e. partial trace over any subsystem yields maximally mixed state.} and as a such it can be used as an entanglement witness\cite{horodecki2001separability}. It means that the operator
\begin{equation}
X=\left(\ket{\psi_+}\bra{\psi_+}\right)^{T_2}
\end{equation}

detects entanglement the following way: if for any state $\rho$
\begin{equation}
\label{eqn:witn}
\mean{\left(\ket{\psi_+}\bra{\psi_+}\right)^{T_2}  }{\rho}<0,
\end{equation} we may infer that $\rho$ is entangled. This works similarly if we consider $\ket{\psi_+}$ transformed by any local unitary $U\otimes\mathbb{1}$:
\begin{equation}
X_U=\left(U\otimes\mathbb{1}\ket{\psi_+}\bra{\psi_+}U^\dagger\otimes\mathbb{1}\right)^{T_2}.
\end{equation}
When looking at the separable numerical range of $X,X_U$ we immediately see that there exist quantum states which entanglement is detected by the separable numerical range, while not by the entanglement witness \eqref{eqn:witn}.
\begin{marginfigure}
\includegraphics{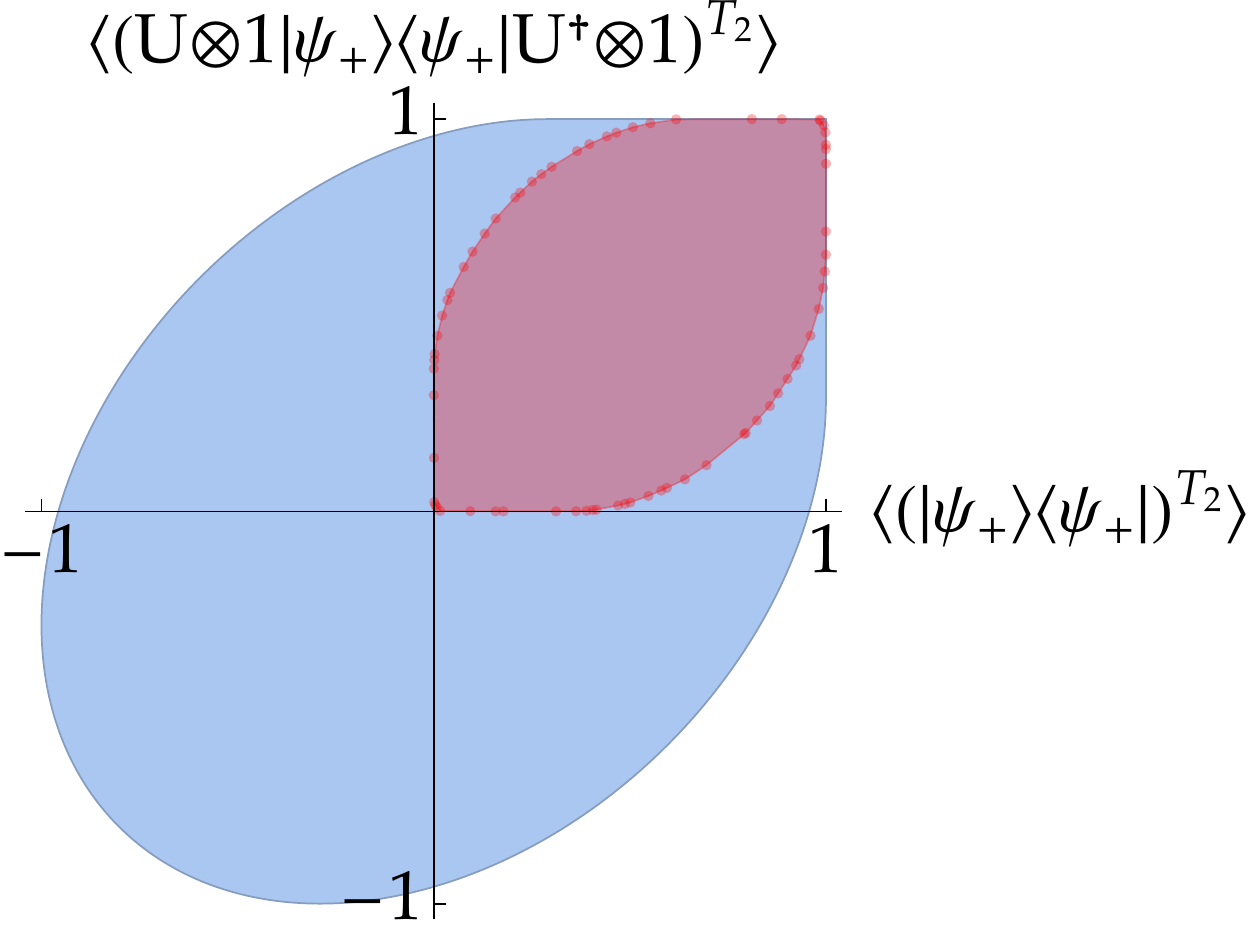}
\caption{Joint numerical range (blue) and separable numerical range (red) of $X,X_U$ for a fixed $U$. If results of a double measurement yield a point in the blue set, existence of entanglement is confirmed. The original linear entanglement witnesses capture only the part with $\mean{X}{}<0$ and $\mean{X_U}{}<0$.}
\end{marginfigure}

\subsection{Thermal Range}
Relation of the points on the boundary of JNR to the ground states of a parametrized Hamiltonians family suggests a natural generalization: as JNR `measures' the ground states, which we may view as the states of the system at zero temperature, we may consider the system at nonzero temperature. So, if we view the function 
\begin{equation}
b(\vec n)=(\mean{F_1}{\ket{\psi}},\ldots,\mean{F_k}{\ket{\psi}})_{\ket{\psi} \text{-- ground state of} \vec n \cdot \vec F}
\end{equation}
returning the point in the boundary of JNR (which maps over $S^{n-1}$ to obtain the whole boundary), we may as well extend it to

\begin{equation}
b_\beta(\vec n)=(\mean{F_1}{\rho},\ldots,\mean{F_k}{\rho})_{\rho \text{-- thermal state of }\vec n \cdot \vec F,~\exp(-\beta \vec n\cdot \vec F)/Z}.
\end{equation}

There are substantial changes between such defined object and the Joint Numerical Range: if we take fixed $\vec n$ and calculate $b_\beta(\vec n)$, generically it is not the point that supports plane with normal $\vec n$. Definition is still consistent, but the resulting object is not as useful if we do not adopt some changes. As a basic requirements, for a given $\vec n$ we should be able to extract from the object
\begin{marginfigure}
\includegraphics{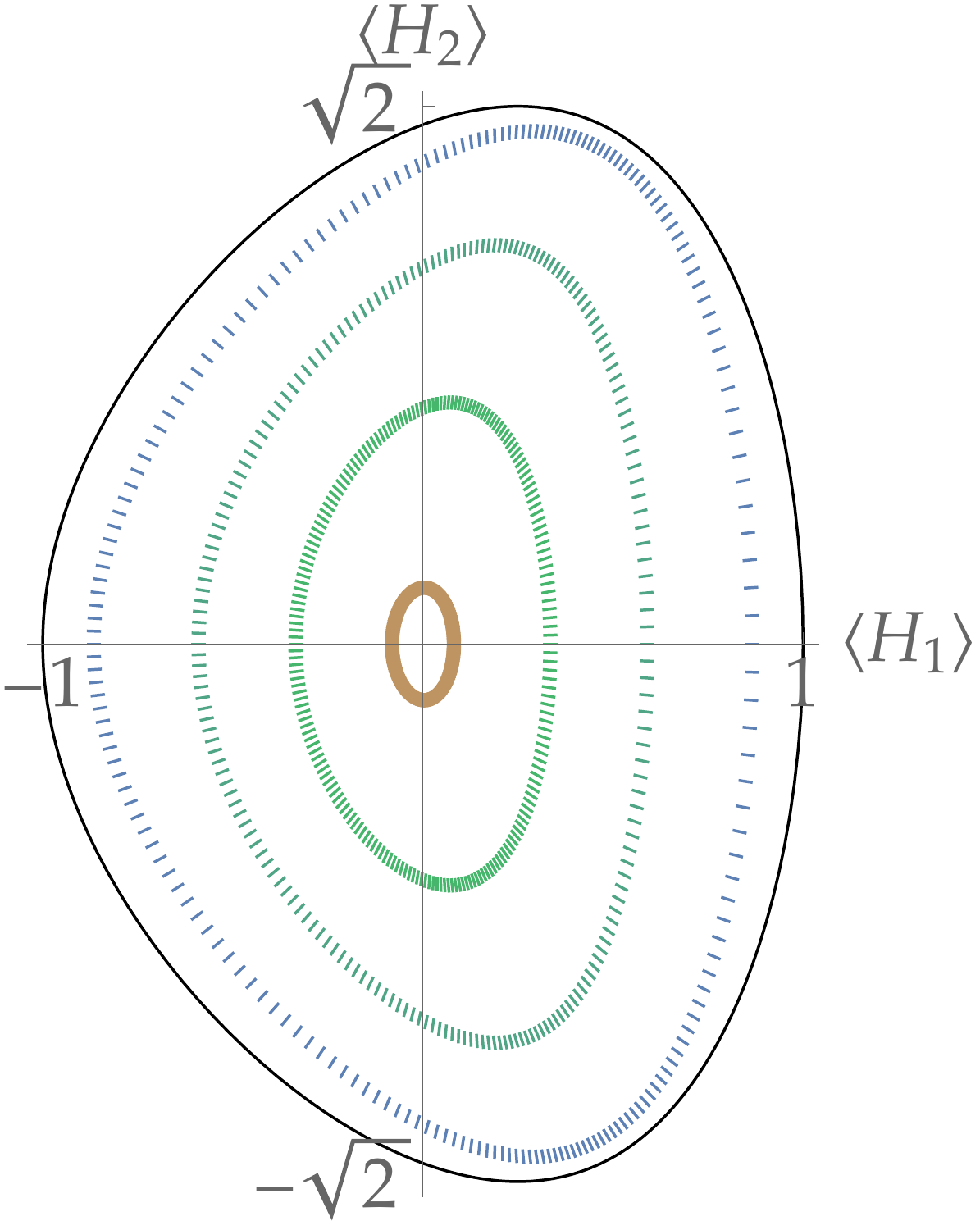}
\caption{Thermal Ranges for different thermodynamic temperatures: $\beta=\infty$ $, 2, 1, 0.5, 0.1$. The smaller the object is, the lower $\beta$. Thermal Numerical Ranges (except for $\beta=\infty$, which coincides with the standard JNR) are shown as the points with attached `fake' normals, so the objects appears hairy. The fact that they do not coincide with real normals is clearly visible. 
}
\end{marginfigure}

\begin{itemize}
	\item information about the energy of a thermal state of Hamiltonian $\vec n \cdot \vec H$,
	\item set of all possible attainable observables of the thermal state.
\end{itemize}
This is easily done: we just need attach in each point the `original' normal $\vec n$ that was passed to the $b_\beta$. This corresponds to selection of fiber bundle section on $k-1$-dimensional surface and as such, has some interesting properties, for instance, the \emph{fake} normal must coincide with \emph{true} normal on $k-2$-dimensional subset of the boundary, plus possibly some isolated points.




\chapter{Phase transitions}
\label{ch:3}
\begin{fullwidth}
This chapter covers the topic of application of the joint numerical range of suitably chosen observables to the theory of phase transitions in many-body systems. In the first place I show the connection between a boundary of the JNR and the ground states of a certain, linearly parametrised, Hamiltonians family. The relation allows one to identify the points of phase transitions and a nonanalytical point at the boundary of JNR. I present examples of physical systems, in which phase transitions can be detected with help of the notion of JNR.
\end{fullwidth}
\section{Connection between JNR and phase transitions}
A quick glance at the algorithm generating the boundary of Joint Numerical Range gives an idea how the JNR formalism can be applied in the theory of phase transitions. The application stems from the property of exposed faces: the one, supporting hyperplane with normal $\vec{n}$, is an image of the projectors onto eigenspace convex sums of these projectors\cite{spitkovsky2017new}.

These states are precisely the ground states of the Hamiltonian $\vec{n}\cdot\vec{F}$ --- by examining the exposed face we can study the properties of the ground state space of the system. 
\section{Simple examples}
A simple nontrivial example is given by the JNR of two operators of size $d=3$ --- as the JNRs operators of size 2 are images of Bloch ball, they are everywhere analytical, except for the case of linear dependence of observables. JNR of two operators of order three is known to display examples with nonanalytical boundary.
\subsection{First order transition}
To demonstrate an example depicting sudden change of the ground state as a function of a parameter, consider Hamiltonian with two terms $H=\alpha H_0+\beta H_1$ with both in the joint block diagonal form,
\begin{marginfigure}
\label{fig:ch2fst}
\includegraphics{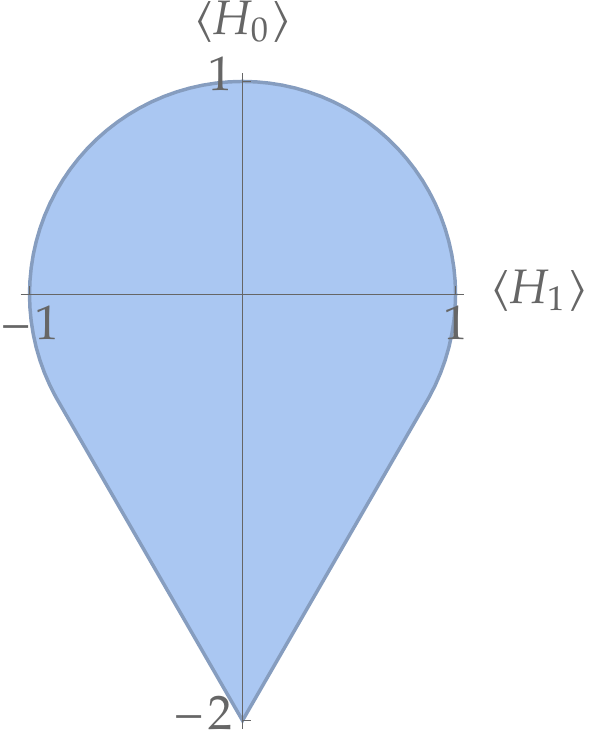}
\caption{Joint numerical range of operators $H_0, H_1$ described by Eq. \eqref{eq:ch2fst}.}
\end{marginfigure}
\begin{equation}
\label{eq:ch2fst}
H_0=\begin{pmatrix}0&1&0\\1&0&0\\0&0&-2\end{pmatrix},~
H_1=\begin{pmatrix}1&0&0\\0&-1&0\\0&0&0\end{pmatrix}.
\end{equation}
The corresponding JNR is shown in Fig. \ref{fig:ch2fst}. Nonanalytical part of the boundary -- a cusp at the negative values of $\mean{H_0}{}$ -- is clearly visible. 
To show the connection between JNR and the behavior of the ground state we define a parametrization of the Hamiltonian
\begin{equation}
\label{eq:ch2fstpa}
H(\theta)=\cos\theta H_0 + \sin\theta H_1
\end{equation}
and plot the eigenvalues as a function of $\theta$. Since we are mostly interested in the ground states, the general linear combination $w H_0 + t H_1$ can be rescaled to obtain this form. Sudden change of the ground state energy at $ \theta=\pm \pi/4 $ is clearly visible --- the reason behind this nonanalytical behavior is evident: it comes from choosing the minimal eigenvalue\footnote{And $\min$ function is not analytical.}.
\begin{figure}
\includegraphics{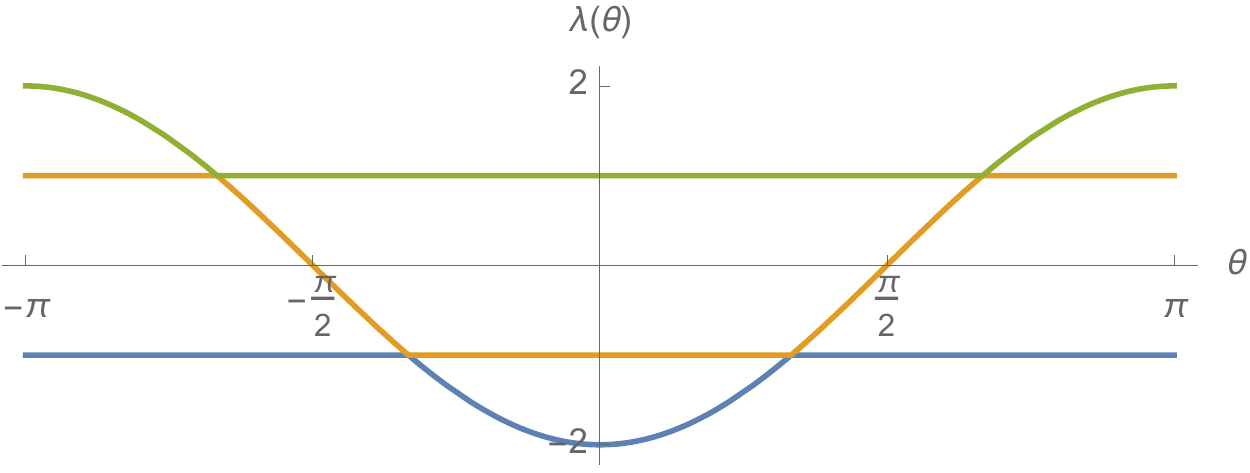}
\caption{Eigenvalues of the Hamiltonian \eqref{eq:ch2fstpa} as a function of phase $\theta$. Note that the angle $\theta_0$, for which an eigenvalue crossing occurs, is visible in the JNR of $H_0, H_1$.}
\end{figure}
\subsection{Second order transition}
\begin{marginfigure}
\includegraphics{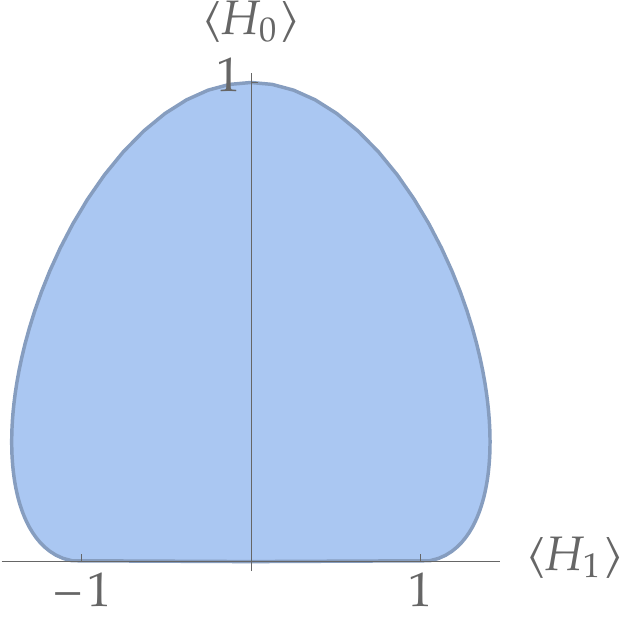}
\caption{Joint numerical range of operators described by Eq. \eqref{eq:ch2snd}.}
\end{marginfigure}
Another simple example, a JNR of two operators
\begin{equation}
\label{eq:ch2snd}
H_0=\begin{pmatrix}0&0&0\\0&0&0\\0&0&1\end{pmatrix},~
H_1=\begin{pmatrix}0&1&0\\1&0&1\\0&1&0\end{pmatrix}
\end{equation}
produces an object with different properties on the boundary. It has no cusps -- instead, a single horizontal face is visible.

\begin{figure}
\includegraphics{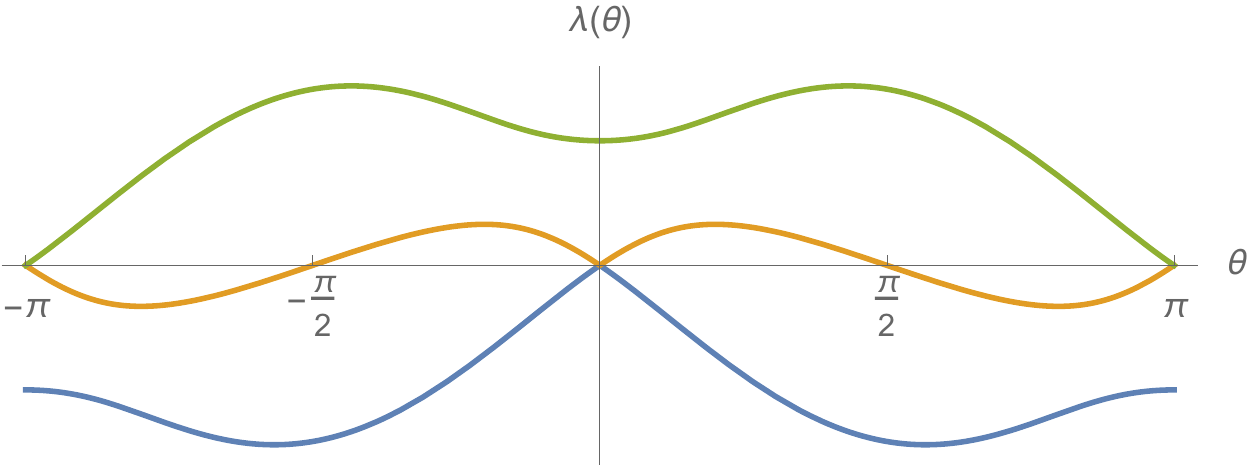}
\caption{Joint numerical range of operators described by Eq. \eqref{eq:ch2snd}.}
\end{figure}
\section{Two interacting particles}
After presenting a link between the analytical properties of the boundary of JNR of two operators $H_1, H_2$ and phase transitions occurring in the system described by the Hamiltonian $H=H_1+y H_2$ under variation of the parameter $y$, we may present a simple physical system, in which the nontrivial change of the ground state space is visible.

It is the case of two qubit system; it may model the two interacting spins, which suggests a natural choice of possible interaction Hamiltonians:
\begin{itemize}
\item Since the classical dipole-dipole interaction has the form\\
 $H_{dd}=|r|^{-3} \left(3\left(\vec s_1 \cdot \frac{\vec r}{|r|}\right)\left(\vec s_2 \cdot \frac{\vec r}{|r|}\right)-\vec s_1 \cdot \vec s_2\right)$, we can choose \\
$3\sigma_x\otimes\sigma_x-\sum\sigma_i\otimes\sigma_i$,
\item interaction with external fields: $\sigma_x\otimes 1+1\otimes \sigma_x$,$\sigma_z\otimes 1+1\otimes \sigma_z$,
\item additional self-interaction: $(\sigma_z\otimes 1+1\otimes \sigma_z)^2$,~\ldots
\end{itemize}
Such a limited choice already provides new, interesting examples to the JNR menagerié. Resulting figures are presented below:

\begin{figure}
\includegraphics{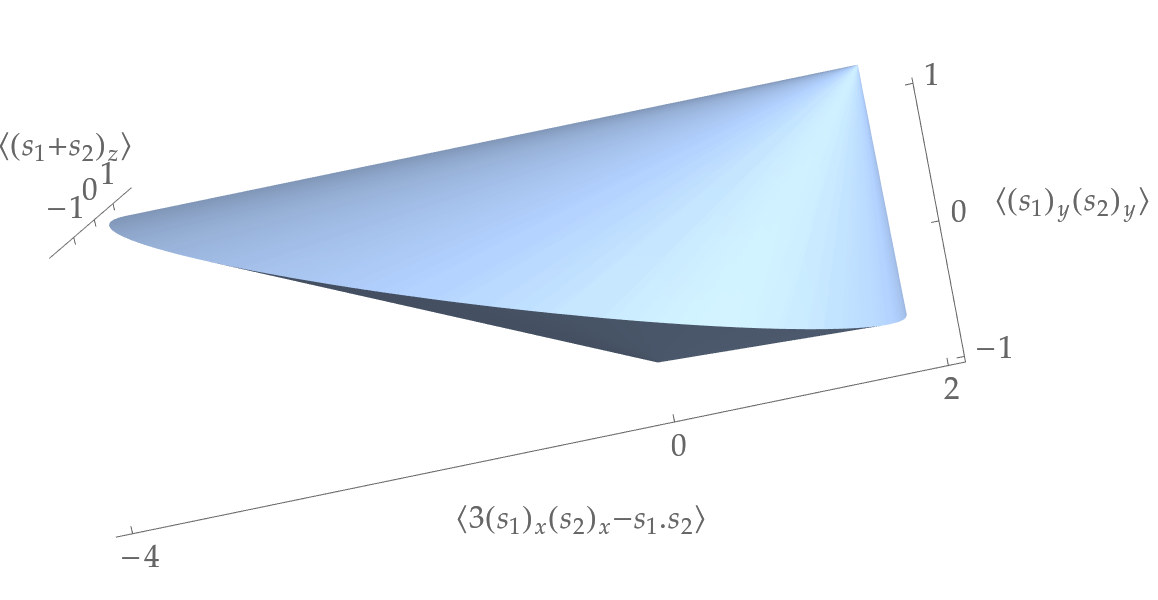}
\caption{
Joint numerical range -- a bicone -- of operators 
\normalsize

$H_1=3\sigma_x\otimes\sigma_x-\sum_{i=1}^3 \sigma_i\otimes\sigma_i,$

$H_2=\sigma_z\otimes 1+1\otimes \sigma_z,$

$H_3=\sigma_y\otimes\sigma_y.$
 }
\end{figure}
\begin{figure}
\includegraphics{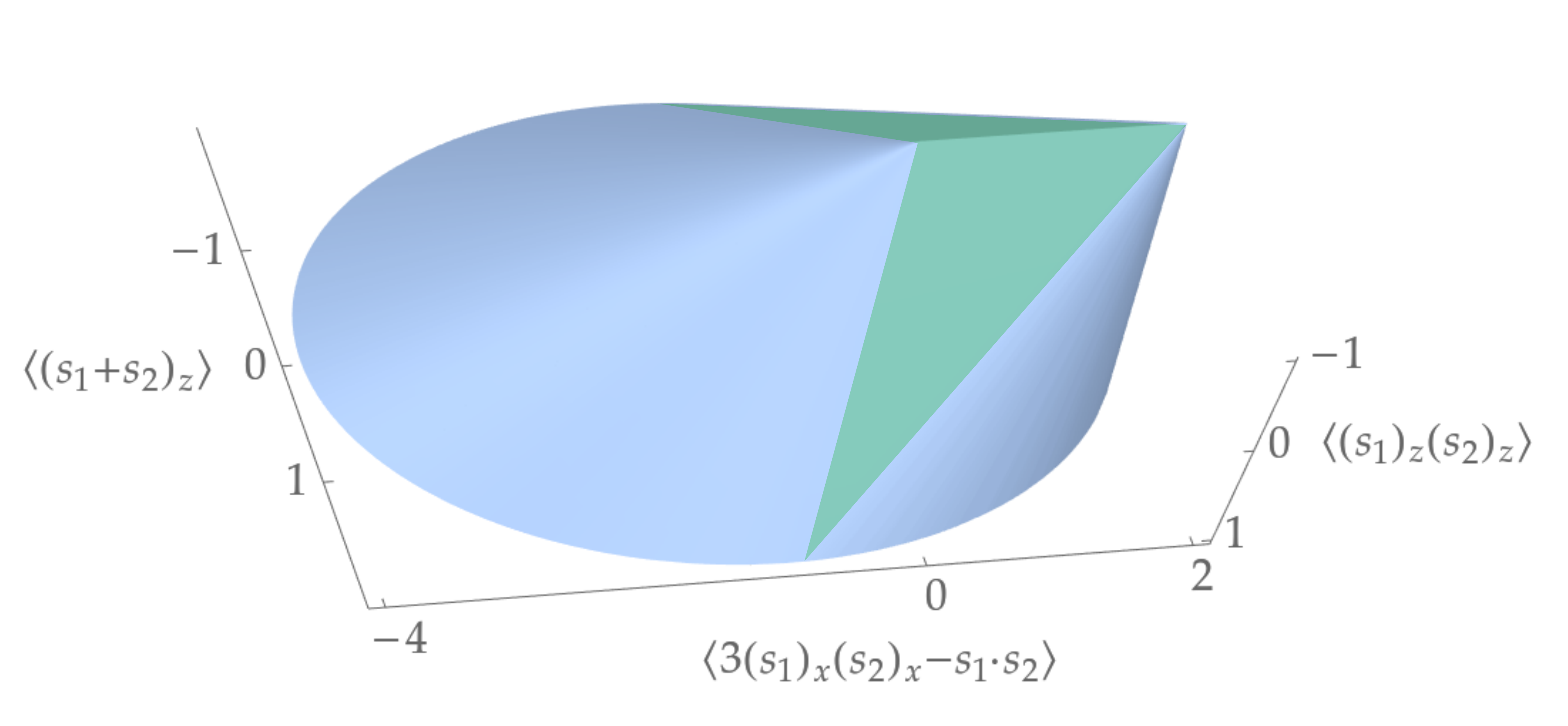}
\caption{Joint numerical range -- a convex hull of an ellipse and a segment -- of three operators of size $d=4$,
{\normalsize

$H_1=3\sigma_x\otimes\sigma_x-\sum_{i=1}^3\sigma_i \otimes\sigma_i,$

$H_2=\sigma_z\otimes 1+1\otimes \sigma_z,$

$H_3=\sigma_z\otimes\sigma_z$.}

There exist two triangles (shown in green), one ellipse and one segment in the boundary.}
\end{figure}

\section{Spin chains}
As a practical application we present JNRs related to several spin chain models. For the sake of simplicity, we consider model where each site has internal qubit states $\ket{0},\ket{1}$, so the total Hilbert space is spanned by the "computational basis"
\begin{equation*}
\begin{split}
\ket{0,0,0,\ldots,0},\\
\ket{0,0,0,\ldots,1},\\
\ldots~~~~~~~~~~\\
\ket{1,1,1,\ldots,1}.\\
\end{split}
\end{equation*}
For convenience let us introduce the following spin operators
\begin{equation}
\begin{split}
s^{(i)}_j &= \underbrace{\mathbb{1}\otimes\mathbb{1}\otimes\ldots}_{i-1\text{ times}}\otimes\sigma_j\otimes\underbrace{\ldots\otimes\mathbb{1}\otimes\mathbb{1}}_{N-i\text{ times}}\\
S_j &= \sum_i s^{(i)}_j.
\end{split}
\end{equation}
Making use of this notation we may introduce operators describing interaction between neighboring sites,
\begin{equation}
\sum_i s^{(i+1 \text{mod} N)}_j s^{(i)}_j,
\end{equation}
or the total spin length operator:
\begin{equation}
S^2=S_x^2 + S_y^2 + S_z^2.
\end{equation}
%
\subsection{Ising model}
Ising model with external fields has the Hamiltonian of form
\begin{equation}
\label{eqn:ising1}
H=\underbrace{-J \sum_i s^{(i+1 \text{ mod } N)}_z s^{(i)}_z}_{\text{spin-spin interaction}} - \overbrace{h \sum_i s^{(i)}_z}^{\text{external transverse field}}- \underbrace{\alpha \sum_i s^{(i)}_{x}}_{\text{external parallel field}}.
\end{equation}
I have explicitly chosen external fields along $z$ and $x$ axes, as it is always possible by rotating the system to express the Hamiltonian in such terms.
If we write the Eq. \eqref{eqn:ising1} in the form
\begin{equation}
\label{eqn:ti1}
\begin{split}
H&=-J N H_1 - h N H_2- \alpha N H_3, \\
H_1&=\frac{1}{N}\sum_i s^{(i+1 \text{ mod } N)}_1 s^{(i)}_1,\\
H_2&=\frac{1}{N}\sum_i s^{(i)}_1,\\
H_3&=\frac{1}{N}\sum_i s^{(i)}_3,
\end{split}
\end{equation}
we immediately see the connection to the JNR: a point in $L(H_1,H_2,H_3)$, supporting plane with normal $(J,h,\alpha)$ is the image of the ground state of the Hamiltonian $H$ with these parameters. The normalization $1/N$ is convenient, as the eigenvalues support of $H_1,H_2,H_3$ is the same for every number of spins $N$.
\begin{figure}
\includegraphics{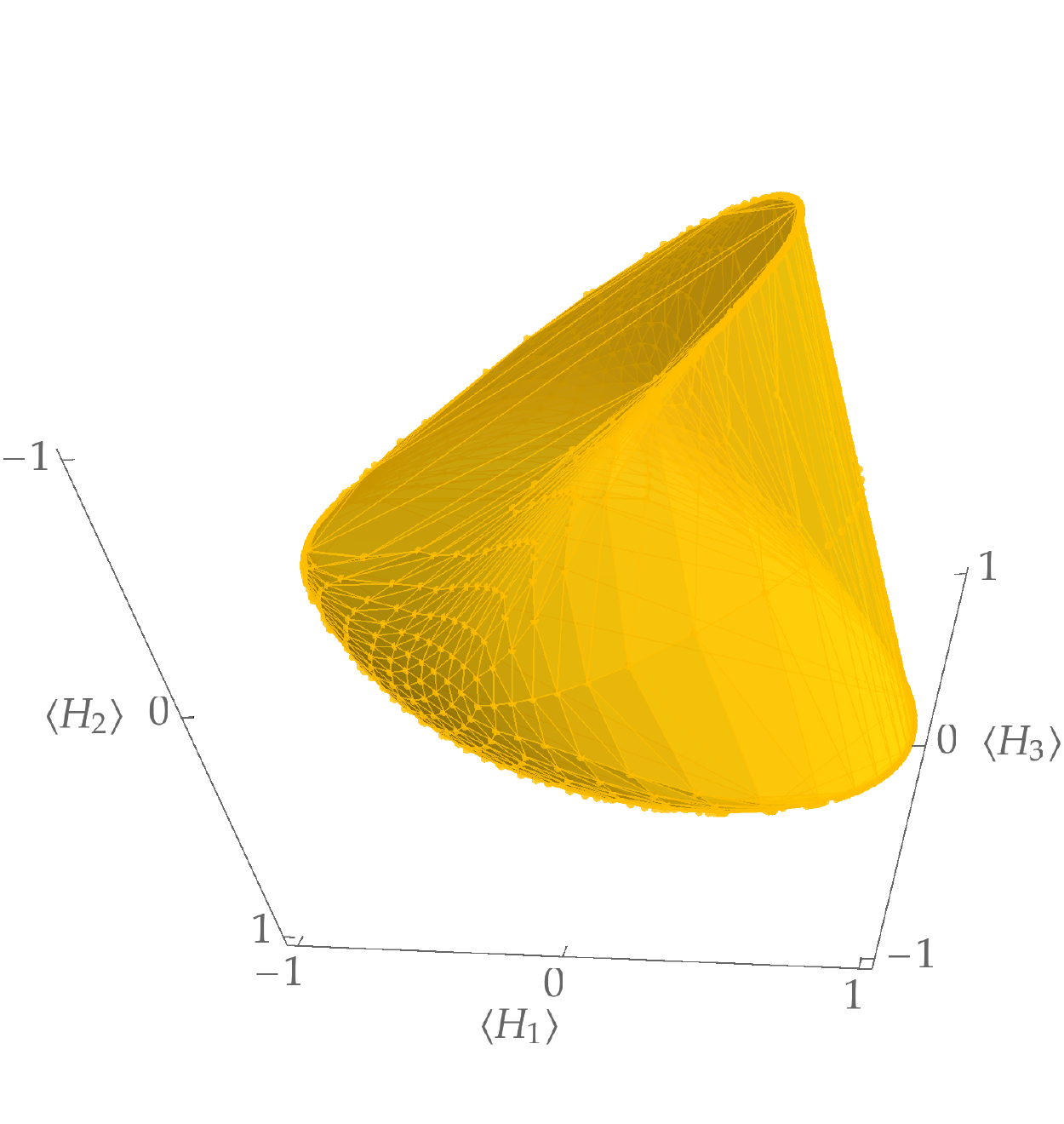}
\caption{Joint numerical range of operators described by Eq. \eqref{eqn:ti1}. In the boundary it is possible to recognize parts relating to the ferromagnetic and paramagnetic behavior as well as sudden ground state changes. In this image the number of sites $N=8$, but increasing the $N$ does not change the figure considerably.}
\end{figure}
~\\~\\~\\~\\~\\
\subsection{Other interactions}
The Ising model  does not show any further interesting behaviors in the boundary and only the trivial phase transitions are visible. Here I present a more intricate three-terms model with the following sub-Hamiltonians
\begin{equation}
\label{eqn:t1}
\begin{split}
H_1&=\frac{1}{N}\sum_i s^{(i+1 \text{ mod } N)}_x s^{(i)}_x,\\
H_1&=\frac{1}{N}\sum_i s^{(i+1 \text{ mod } N)}_z s^{(i)}_z,\\
H_3&=\frac{1}{N(N+1)}\left((S_x)^2+(S_y)^2+(S_z)^2\right).
\end{split}
\end{equation}
As before, the normalization has been set to ensure equal eigenvalues supports.
The resulting figures possess interesting features of the boundary, including multiple curved surfaces, connecting nontrivially to the flat parts.

Such behavior of the flat parts is not only of mathematical interest -- as we have seen before, each flat part indicates a presence of a phase transition. The existence of ruled surfaces means that the ground state of certain Hamiltonian must break symmetry -- here we see that it breaks the symmetry in a nontrivial fashion, unlike the case of Ising model. If multiple flat parts are present, it is possible to choose the final symmetry breaking ground state by choosing the path of Hamiltonians. In general, notion of JNR presents the properties of interesting Hamiltonians better than the usual approaches, even if the final information content is the same.
\begin{figure}
\includegraphics{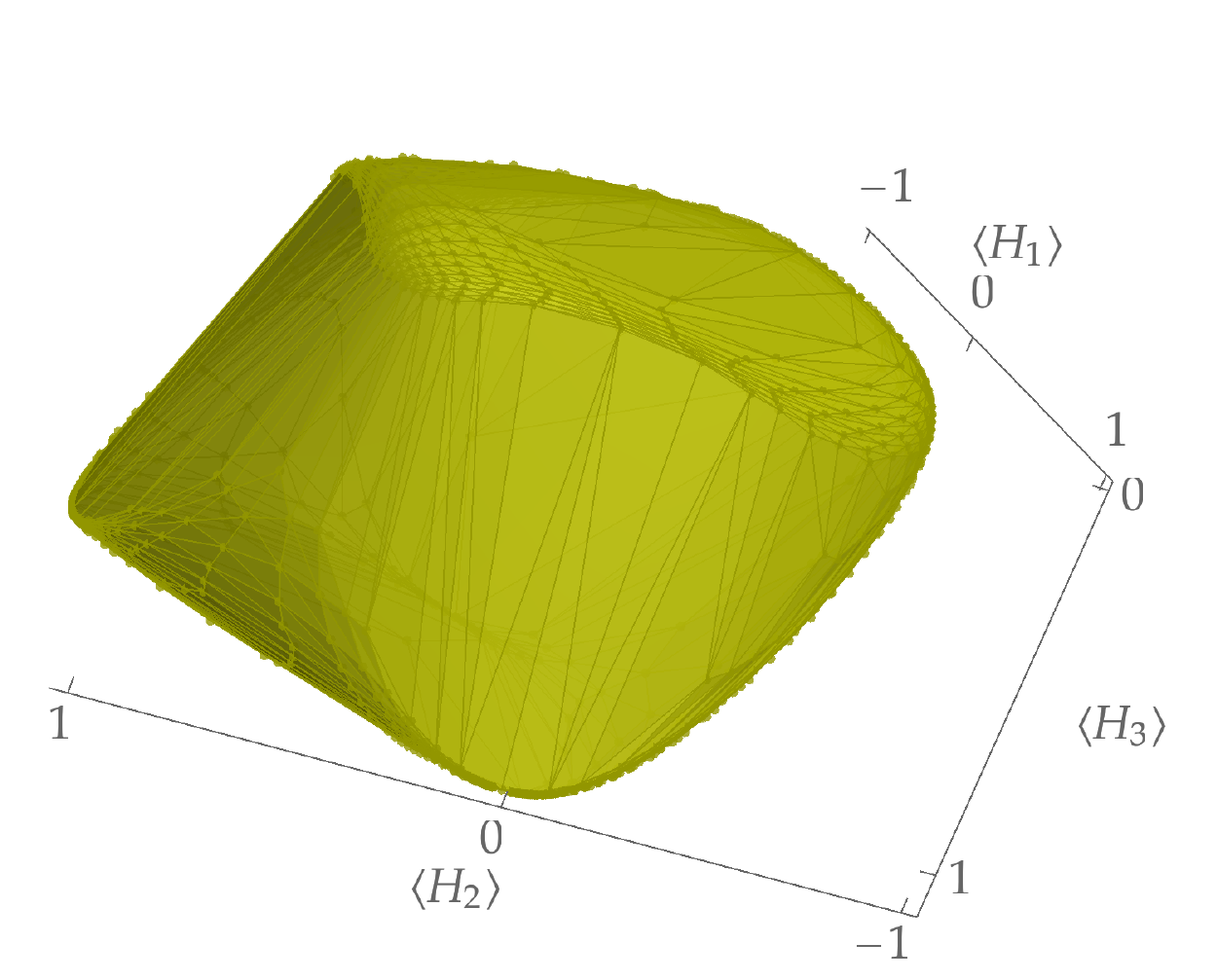}
\includegraphics{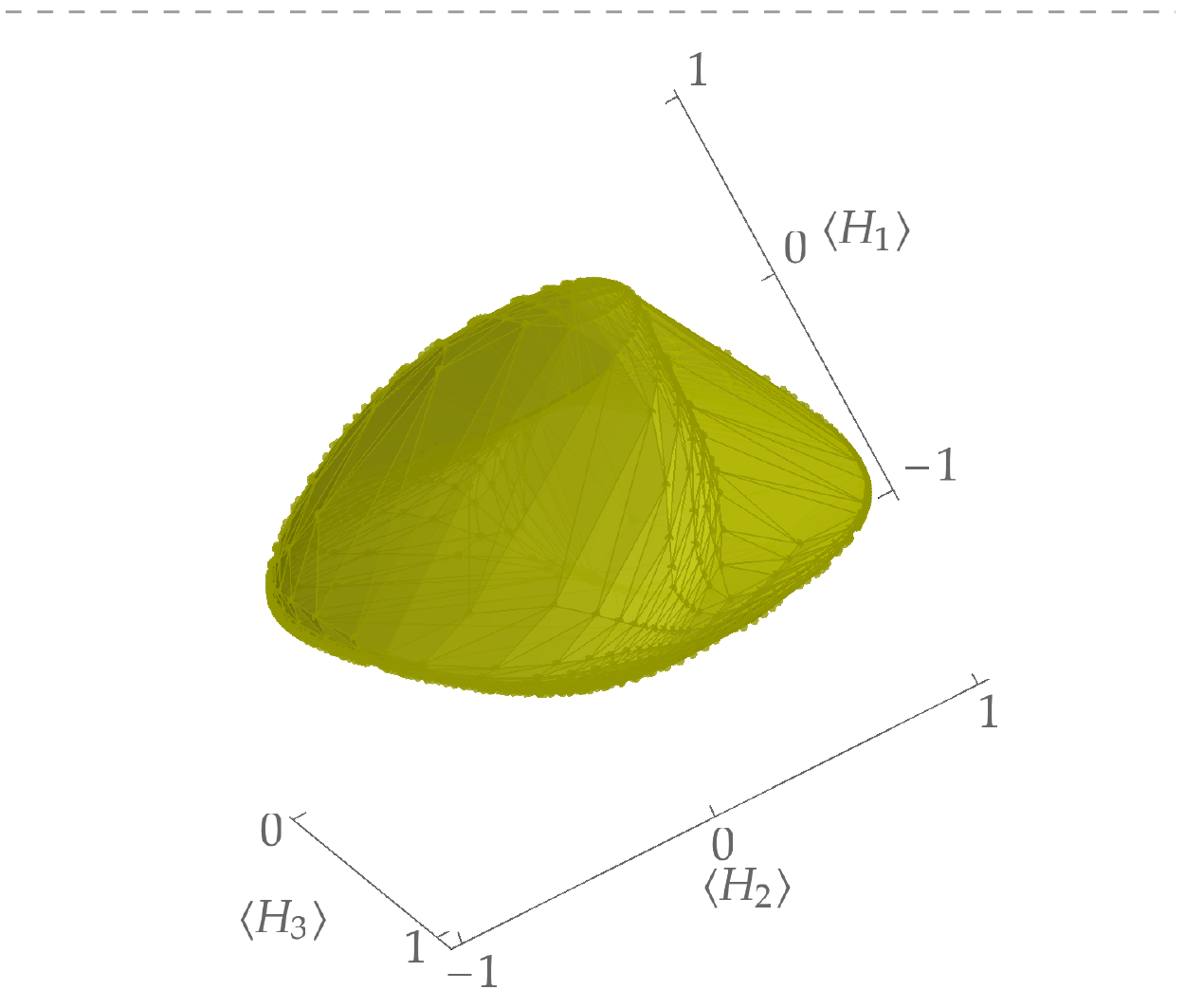}
\caption{Joint numerical range of operators described by Eq. \eqref{eqn:t1} viewed from several different angles. In this image the number of sites $N=10$ saturates the JNR.}
\end{figure}
\clearpage
\section{Bounds for the ground state energy}
\begin{marginfigure}
\includegraphics{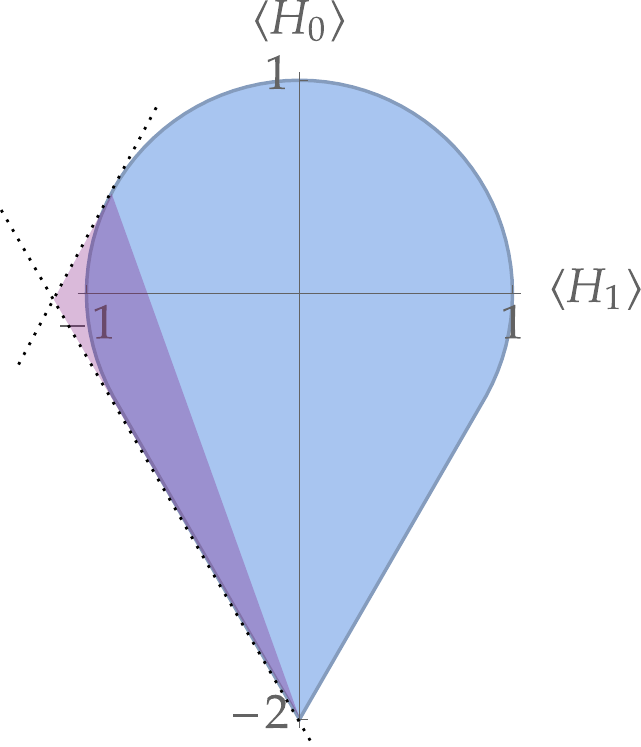}
\caption{Visual depiction of the meaning of the bound. If two point on the boundary (along with their normal vectors) are known, we may infer that arc of the JNR boundary is contained in the shaded region because of the object convexity, which restricts the choices of ground state energies in some parameters region.}%
\end{marginfigure}
Based on the JNR it is possible to provide bounds for the ground state energies of a linearly parametrized family of Hamiltonians. Consider the following problem: given parametrization
\begin{equation}
H(a) = H_0 + a H_1
\end{equation}

and ground state energies of, say $H(0)$ and $H(1)$, can we provide a bound for the ground state energy in other parameter range? It turns out that the answer is positive --- as the ground state energy is exactly the \emph{support function} of JNR, and JNR is convex, the ground state energy as a function of $a$ is \emph{concave}. This means -- from the very definition of concavity -- that if we know the values of $H(0)$ and $H(1)$, for $a\in[0,1]$
\begin{equation}
H(x)\ge a H(0)+(1-a)H(1).
\end{equation}
~\\~\\~\\~\\~\\~\\
\begin{figure}
\includegraphics{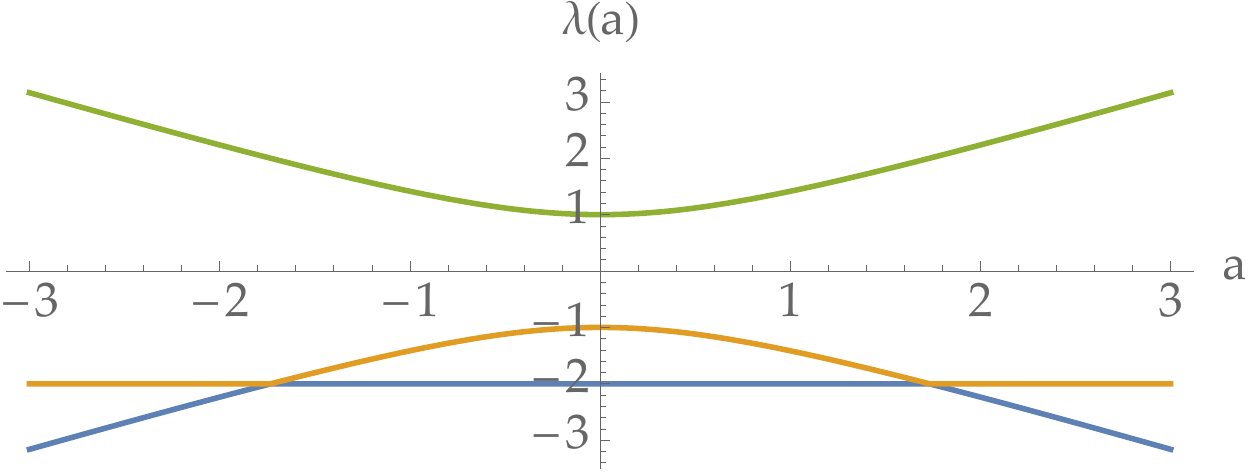}

\caption{Spectrum of $H_0+a H_1$. The concavity of ground state energy as a function of the parameter $a$ is clearly visible.}
\end{figure}



\chapter{Uncertainty relations}
\begin{fullwidth}
In this chapter, a connection between JNR and some forms of the finite-dimensional uncertainty relations is shown, along with practical algorithms for determination of the bounds.
\end{fullwidth}
\section{General uncertainty relation}
Several known uncertainty relations based on variances, the Heisenberg-Robertson-Schrödinger\cite{angelow1999heisenberg}
\begin{equation}
(\Delta^2 X)(\Delta^2 Y) \ge c_\times(X,Y),
\end{equation}
or Maccone-Pati\cite{maccone2014stronger}
\begin{equation}
\Delta^2 X + \Delta^2 Y \ge c_+(X,Y), 
\end{equation}
have form which suggests a natural generalization to the form of bounds for an \emph{uncertainty function},
\begin{equation}
\label{eq:uncerfun}
u(\Delta^2 F_1,\ldots,\Delta^2 F_k)\ge c_u (F_1,\ldots,F_k),
\end{equation}
which is minimized over the set of quantum states $\mathcal{M}_d$. Reasonable definition of such a function $u$ should contain one constraint: $u$ should be minimized on pure states, as minimization of \emph{uncertainty} over a \emph{mixture} is a counterintuitive behavior.                   
%
%
%
%
\section{Relation to JNR}
The connection between JNR and general variance-based uncertainty relation is straightforward: the function of form \eqref{eq:uncerfun} may be viewed as acting on the Joint Numerical Range of certain operators 
\begin{equation}
L(F_1,F_1^2,F_2,F_2^2,\ldots,F_k,F_k^2)
\end{equation}
transformed under the map
\begin{equation}
\Delta: (\overbrace{f_1}^{\mean{F_1}{}},\underbrace{f'_1}_{\mean{F_1^2}{}},\ldots,\overbrace{f_k}^{\mean{F_k}{}},\underbrace{f'_k}_{\mean{F_k^2}{}}) \mapsto (\overbrace{f'_1-f_1^2}^{\Delta^2 F_1},\ldots,\overbrace{f'_k-f_k^2}^{\Delta^2 F_k})
\end{equation}
One of the possible realizations of the condition of $u$ (being minimized on pure states) is relatively simple and useful\footnote{It is not the only possible realizations though.}: if the compound function $u\circ\Delta$ is \emph{concave}\footnote{$\circ$ denotes function composition.}, it is minimized on the boundary of $L$. This condition is realized in two simple cases, $u_+$ and $u_\times$.\footnote{Note that $u_\times$ alone is \emph{not} a concave function, while $u_\times \circ \Delta$ is.}.
\section{Numerical analysis}
Concave uncertainty functions are minimized on the boundary of corresponding Joint Numerical Range. This fact allows us to determine numerical approximations for the lower bounds of uncertainty, $c$. In fact, due to convex structure it is possible to provide upper and lower bounds for $c$ -- since we know a discrete subset of the point on the boundary $ \bar\delta L $ along with the discrete subset of the  supporting hyperplanes $\bar H$, we can provide two approximations to JNR:
\begin{marginfigure}
\includegraphics{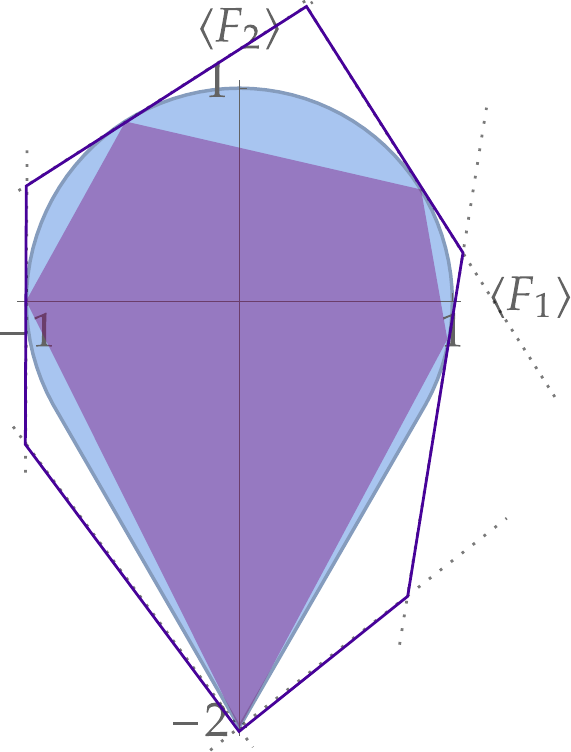}
\caption{Internal and external approximating polygons of a certain JNR. The inner polygon is convex hull of sampled points on the boundary, the external one is generated as an intersection of supporting half-spaces.}
\end{marginfigure}
\begin{enumerate}
	\item a subset of true JNR, convex hull of the generated points,
	\item a superset of JNR, intersection of the half-spaces generated by supporting hyperplanes. This generates a polytope -- different from the previous one.
\end{enumerate}
From each approximation we determine different bound for the uncertainty relation -- since the uncertainty function was assumed to be convex, it is sufficient to check the the corner points of polytopes only\footnote{In the case of external approximating polytope it is suboptimal, but sufficient and simpler.}.
This method provides two significant advantages over sampling the whole state space to find the lower bound
\begin{itemize}
\item we try to minimize a function on $S^{2k-1}$ (real sphere of normal vectors), which in most cases has smaller dimension than state space (complex $S^{2d-1}$ sphere),
\item we obtain both approximations from above and below for the true bound simultaneously.
\end{itemize}
\subsection{Sum of two variances}
Consider the Maccone-Pati uncertainty relation for the sum of variances -- our aim is to provide the lower bound for
\begin{equation}
\label{eqn:parabo}
u(\Delta^2 X,\Delta^2 Y)=\Delta^2 X+\Delta^2 Y.
\end{equation}
This case allows for a simplified analysis: as $\Delta^2 X+\Delta^2 Y=\mean{X^2+Y^2}{}-\mean{X}{}^2-\mean{Y}{}^2$, there is no need for considering a four-dimensional object $L(X,X^2,Y,Y^2)$. A simple, 3-dimensional joint numerical range $L(X,Y,X^2+Y^2)$  is sufficient to investigate bounds for the sum of variances -- the surfaces of equal uncertainty are paraboloids:
\begin{equation}
	\mean{X^2+Y^2}{}-\mean{X}{}^2-\mean{Y}{}^2=c.
\end{equation}

In the case of small size of matrices $d\le4$ it is possible to analyze the uncertainty relation analytically. In short -- it is possible to transform both the JNR boundary and the parabola (Eq. \eqref{eqn:parabo}) to a projective dual space, where calculations are easier. The points where two surfaces `touch' each other\footnote{Intersect and have colinear normal vectors.} in this space are the candidates for places of uncertainty minimization. 

\subsection{Qutrit examples}
To illustrate the role of JNR in uncertainty relations, we present the simple qutrit examples of Maccone-Pati uncertainty relation analyzed using previously shown methods. We restrict to $d=3$, as when the analyzed operators $X, Y$ have larger size, many faces and nonanalytical points emerge in $L(X,Y,X^2+Y^2)$, which obscures the view. The $d=2$ case generates flat JNRs, as for operators on a qubit $X^2+Y^2$ is a linear combination of $X, Y$ and identity.
The first analyzed pair of operators is
\begin{equation}
\label{eqn:un1}
X=\begin{pmatrix}0&1&0\\1&0&0\\0&0&0\end{pmatrix},~~~~Y=\begin{pmatrix}1&0&0\\0&0&0\\0&0&-1\end{pmatrix}.
\end{equation}
\begin{figure}
\includegraphics{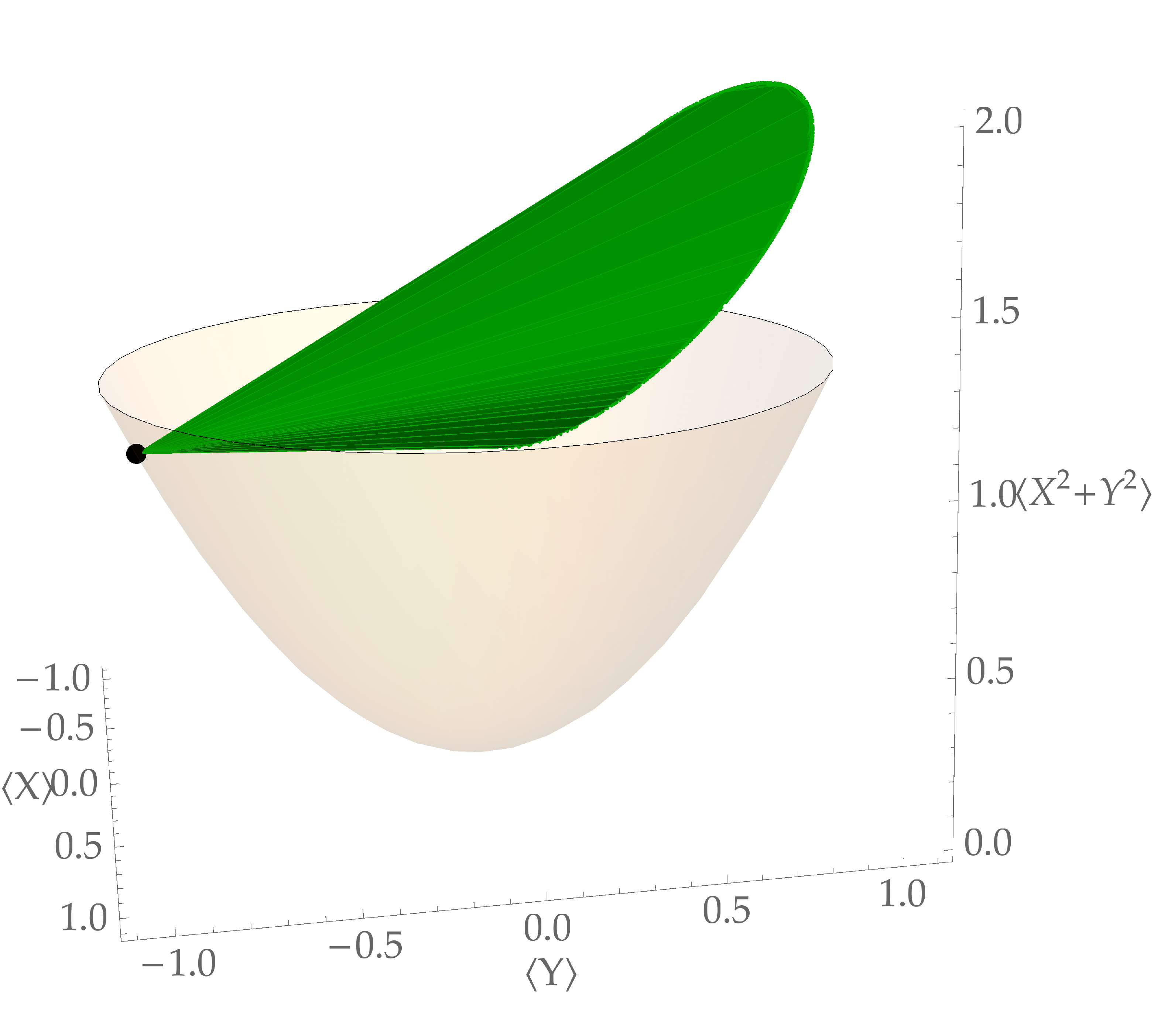}
\caption{Illustration of JNR corresponding to Maccone-Pati uncertainty relation $\Delta^2 X+\delta^2 Y\ge c(X,Y)$ for $X, Y$ described by Eq. \eqref{eqn:un1}. The point on the boundary which minimizes the uncertainty is shown as black dot.
Since the operators have a common eigenvector $(0,0,1)^T$, the uncertainty relation becomes trivial, $\Delta^2 X+\delta^2 Y\ge 0$. This bound is consistent with result of analytical study in dual space and numerical analysis.
}
\end{figure}
\clearpage
The second analyzed pair of operators is
\begin{equation}
\label{eqn:un2}
X=\begin{pmatrix}0&1&0\\1&0&i\\0&-i&0\end{pmatrix},~~~~Y=\begin{pmatrix}1&0&0\\0&0&0\\0&0&-1\end{pmatrix},
\end{equation}
and produces much more interesting result.
\begin{figure}
\includegraphics{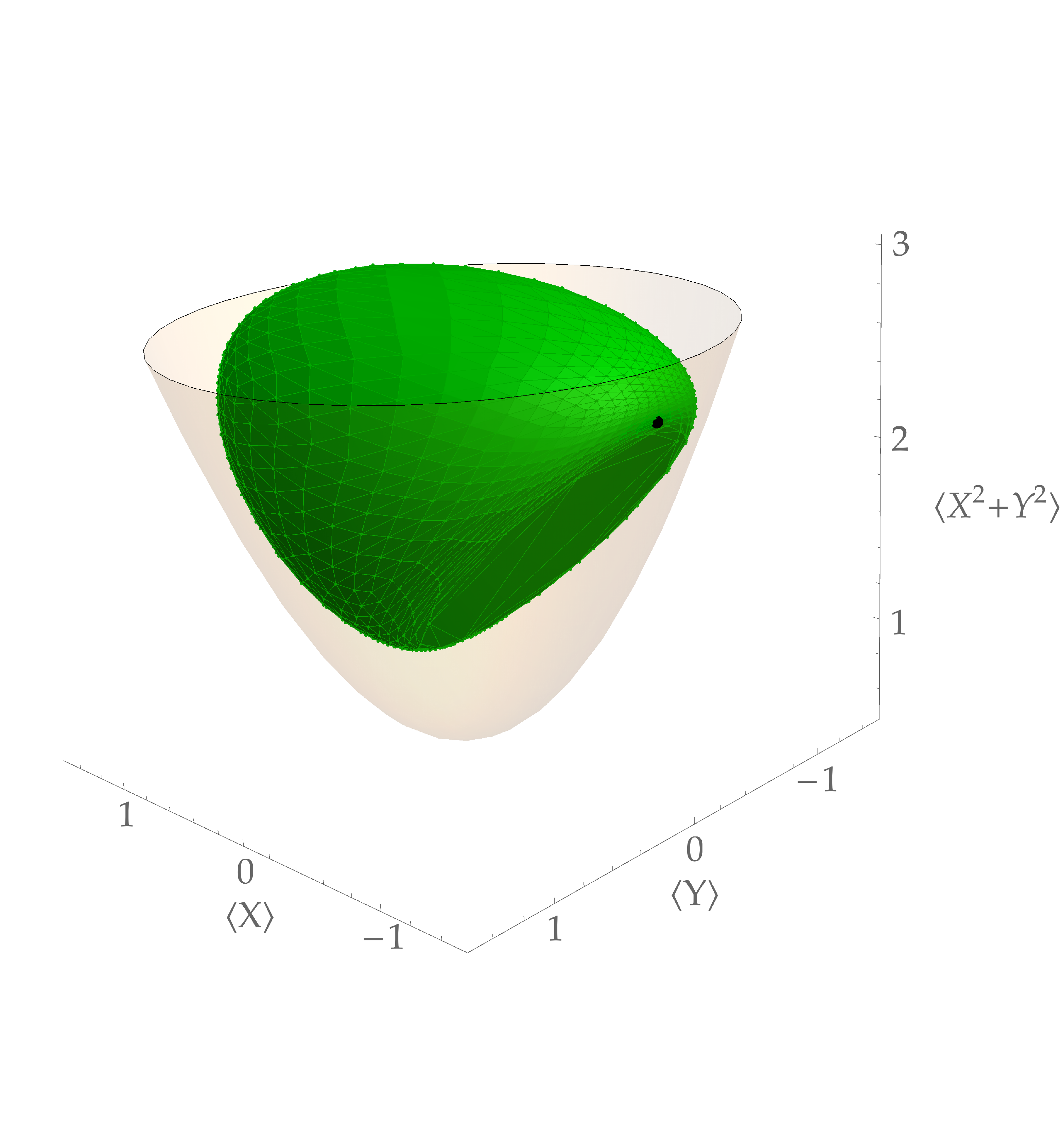}
\caption{Illustration of JNR corresponding to Maccone-Pati uncertainty relation $\Delta^2 X+\delta^2 Y\ge c(X,Y)$ for $X, Y$ described by Eq. \eqref{eqn:un2}. The point on the boundary which minimizes the uncertainty is shown as black dot.
The operators $X, Y$ do not share any common eigenvector. This produces the Joint Numerical Range classified as $s=0, e=2$ -- $L(X,Y,X^2+Y^2)$ has two ellipses in the boundary and no segment is present. 
The uncertainty is minimized on an analytical part of the boundary.
}
\end{figure}

The analytical bound for this case is $c_+ = \frac{15}{32}$. It is consistent with numerical analysis: for 1082 points generated on the boundary, the lower bound from inner approximating polytope is $c_+ - 8.8\times 10^{-4}$, the upper bound from outer approximating polygon is %
$c_+ + 1.2\times 10^{-3}.$

\chapter{Conclusions and open questions}
In this thesis, we have presented the theory of Joint Numerical ranges and its possible applications to analysis of phase transitions and uncertainty relations. Primary result of this work are
\begin{itemize}
	\item introduction of thermal range,
	\item new examples of JNR corresponding to real-life Hamiltonians,
	\item proof of convexity of ground state energy of a linearly parametrized family of Hamiltonians,
	\item numerical, bound-error method for the determination of uncertainty bound,
	\item analytical method for the determination of bound for Maccone-Pati uncertainty relation for low dimensional cases.
\end{itemize}
Some questions still remain open.
\begin{itemize}
	\item Does the existence of vector field on the boundary of thermal range provide new insight to phase transitions?
	\item Is it possible to provide better bounds for ground state energy?
	\item Is it possible to provide a chain of classifications of JNRs, just like the classification of a qubit JNRs has been used in qutrit JNR?
\end{itemize}
\nocite{verstraete2006matrix,zauner2016symmetry,zauner2016njp}


\backmatter

\makeatletter
\renewcommand{\@biblabel}[1]{\textit{(#1)}}
\makeatother
\bibliography{bibliography} 
\bibliographystyle{unsrt} 



\end{document}